\documentclass{amsart}

\usepackage{amssymb}
\usepackage{amssymb,amsmath,amscd}
\usepackage{pstricks,pst-node}
\psset{unit=.5cm}

\usepackage{graphicx}
\usepackage{epstopdf}
\usepackage{amsmath, amsthm, amssymb, amsbsy,amsfonts,amscd}
\usepackage[mathscr]{eucal}
\usepackage{epsfig}
\usepackage{young}
\newtheorem{theorem}{Theorem}[section]

\newtheorem{definition}[theorem]{Definition}
\newtheorem{proposition}[theorem]{Proposition}
\newtheorem{lemma}[theorem]{Lemma}

\newtheorem{example}[theorem]{Example}

\newtheorem{remark}[theorem]{Remark}

\usepackage{amsfonts}
\usepackage{xy}
\usepackage{amssymb}
\usepackage{amsmath}

\def\proof{\par{\it Proof}. \ignorespaces}
\def\endproof{{\ \vbox{\hrule\hbox{%
     \vrule height1.3ex\hskip0.8ex\vrule}\hrule }}\par}

\numberwithin{equation}{section}

\numberwithin{figure}{section}

\let\trueint=\int
\let\truesum=\sum
\def\int{\mathop{\textstyle\trueint}\limits}
\def\sum{\mathop{\textstyle\truesum}\limits}

\let\<=\langle
\let\>=\rangle
\def\Real{{\mathbb{R}}}

\renewcommand\labelitemi{\ifmmode\circ\else$\circ$\fi}

\def\Sym{\mathcal{S}}
\def\R{\mathbb{R}}
\def\C{\mathbb{C}}
\def\Gr{{\rm Gr}}

\def\v{\mathbf{v}}
\def\w{\mathbf{w}}
\def\t{\mathbf{t}}
\def\0{\mathbf{0}}
\newlength{\myVSpace}
\newlength{\bigmyVSpace}
\newcommand\bigxstrut{\raisebox{\bigmyVSpace}
  {\rule{0pt}{\bigmyVSpace}}%
}

\newcommand{\QQ}{\mathsf Q}

\DeclareMathOperator{\conv}{Conv}
\DeclareMathOperator{\CH}{Conv}
\DeclareMathOperator{\Perm}{Perm}

\DeclareMathOperator{\M}{\mathcal M}

\begin{document}


\title[Fifty Years of the Finite Nonperiodic Toda Lattice]{Fifty Years of the Finite Nonperiodic Toda Lattice: \\ A Geometric and Topological Viewpoint}

\author{Yuji Kodama$^1$}
\address{Department of Mathematics, Ohio State University, Columbus,
OH 43210}
\email{kodama@math.ohio-state.edu}
\author{Barbara A. Shipman$^{2}$}
\thanks{$^1$Partially
supported by NSF grants DMS-1410267 and DMS-1714770}

\address{Department of Mathematics, The University of Texas at Arlington, Arlington TX}
\email{bshipman@uta.edu}

\begin{abstract} In 1967, Japanese physicist Morikazu Toda published a pair of seminal papers in the Journal of the Physical Society of Japan that
exhibited soliton solutions to a chain of particles with nonlinear interactions between nearest neighbors.
In the fifty years that followed, Toda's system of particles has been generalized in different directions, each with its own analytic, geometric, and topological characteristics.  These are known collectively as the Toda lattice.  This survey recounts and compares the various versions of the finite nonperiodic Toda lattice from the perspective of their geometry and topology.  In particular, we highlight the polytope structure of the solution spaces as viewed through the moment map, and we explain the connection between the real indefinite Toda flows and the integral cohomology of real flag varieties.
\end{abstract}

\maketitle

\thispagestyle{empty}
\pagenumbering{roman}\setcounter{page}{1}
\tableofcontents

\pagenumbering{arabic}
\setcounter{page}{1}

\section{Historical overview}

In 1974, Henon \cite{Henon}
and Flaschka \cite{Flaschka I} announced the complete integrability of the (real, finite, periodic) Toda lattice. 
This came seven years after the pivotal papers of M.~Toda on vibrations in chains with nonlinear interactions \cite{Toda I} and waves in anharmonic lattices \cite{Toda II}.
Again in 1974, Flaschka \cite{Flaschka I, Flaschka II}, and also Manakov \cite{manakov:74}, showed that the periodic Toda lattice can be written in Lax form through a change of variables, so that the constants of motion appear as eigenvalues of the Lax matrix \cite{Lax}.

Six years later, Moser \cite{Moser} showed that the (real, finite) nonperiodic Toda lattice is completely integrable.
In two different expressions of the equations, the flows obey a Lax equation on a set of real tridiagonal Lax matrices with positive subdiagonal entries. 
The matrices are symmetric in one formulation and Hessenberg in the other -- these are two different expressions of the very same system.  
The flows exist for all time and preserve the spectrum of the initial Lax matrix.  

However, when we allow the entries on the subdiagonal to take on any real values, the tridiagonal symmetric and Hessenberg forms create two genuinely different dynamical systems.
 In the symmetric case, the flows exist for all time and the isospectral manifolds are compact \cite{Tomei}, while in the Hessenberg form, the flows can blow up in finite time and the isospectral manifolds are not compact (see e.g. \cite{KY:96, KY:98}). 
Shortly after this work, the flows on full symmetric real matrices in generic case were shown to be completely integrable \cite{DLNT:86} with the introduction of additional constants of motion.

Section \ref{Sec:FiniteNonperiodicReal} describes these versions of the real, finite, nonperiodic Toda lattices with a focus on their geometry and topology.  This represents the work on real Toda lattices during roughly the first twenty years after the Toda lattice was discovered in 1967.  

By the time the Toda lattice had been known for 25 years, studies on nonperiodic complex versions began to appear.
These are the focus of Section \ref{Sec:ComplexTodaLattices}.
This phase brings in the new idea of compactifying the flows through embeddings into flag varieties. 
In a seminal paper, 
Ercolani, Flaschka, and Haine \cite{EFH} describe the Toda system on complex tridiagonal matrices in Hessenberg form.  
The complex flows again blow up in finite (complex) time,
but they differ from the real 
in that they no longer preserve ``signs" of the subdiagonal entries.  
A theorem on matrix factorizations \cite{Kostant} is used to embed the isospectral sets into a flag variety.  
There, the flows enter lower-dimensional cells, called the Bruhat cells in the Bruhat decomposition of the flag variety,  at the blow-up times, where the singularity at a blow-up time is characterized by the Bruhat cell \cite{FH:91a,AvM:91,CK:04}.

Two years after \cite{EFH},  integrability was extended to the full Kostant-Toda lattice in \cite{EFS}, where the system evolves on complex Hessenberg matrices with arbitrary entries everywhere below the diagonal.  On
isospectral sets with distinct eigenvalues, the flows generate a diagonal torus action under the appropriate embedding into a flag variety.
Similar embeddings, derived from the companion and Jordan matrices of the spectrum, are helpful in understanding nongeneric flows where eigenvalues coincide \cite{Nongeneric} and in describing the compactified complex isospectral sets \cite{Complex Tridiagonal}.  Coincidence of eigenvalues is seen in splittings of moment polytopes, which allows for a description of monodromy around nongeneric isospectral sets in special cases \cite{Monodromy, Sl4}.  

Section \ref{Other Extensions} discusses other extensions of the finite nonperiodic Toda lattices. 
The Toda flow in Lax form is introduced on an arbitrary diagonalizable matrix in \cite{general matrix} and is integrated by
inverse scattering (or equivalently, by factorization).
The tridiagonal Hessenberg and symmetric Toda lattices, which are defined on the Lie algebra of type $A$ (that is, $\frak{sl}_n$),
are extended to semisimple Lie algebras using the Lie algebra splittings from the Gauss (or LU-) and QR-factorizations, respectively \cite{CK:02, CK:02a}.
Related hierarchies are the Kac-van Moerbeke system, which can be considered as a 
square root of the Toda lattice \cite{GHSZ:93, KM:75}, and
the Pfaff lattice, which evolves on symplectic matrices, and is connected to the indefinite Toda lattice \cite{AM:02, KP:07, KP:08}.

Section \ref{RealKT} considers the full Kostant-Toda hierarchy in the real variables \cite{KW:15}.
It classifies the \emph{regular} solutions of the hierarchy in terms of
the \emph{totally nonnegative} parts of the flag variety $G/B$ where $G=SL(n,\R)$ and $B$ is the
set of upper-triangular matrices. Using the moment map, the full Kostant-Toda flows are defined on the appropriate
weight space, and it is shown that the closure of each flow forms an interesting convex polytope,
which we call a \emph{Bruhat interval polytope} (see also \cite{TW:15}). This section begins with a brief review
of the totally nonnegative flag variety \cite{MR, KW3}. The goal is to describe the topological structure of the regular solutions of the full Kostant-Toda lattice for the real split algebra $\frak{sl}_n(\R)$.

Section \ref{Toda-cohomology connections} describes how the \emph{singular} structure of blow-ups in solutions of the \emph{indefinite} Toda lattice contains information about the integral cohomology of real flag varieties \cite{CK:06, CK:07}. 
We consider the moment polytope, the image of the moment map of the isospectral variety, for the real split semi-simple Lie algebra of $\frak{sl}_n(\R)$.
The vertices of the polytope are the 
 orbit of the Weyl group action \cite{CK:02a, FH:91}.  These vertices correspond to the fixed points of the Toda flows.
Each edge of the polytope can be considered as an orbit of the $\frak{sl}_2(\R)$ Toda lattice (the smallest
nontrivial lattice).  
An orbit may be regular (without blow-ups) or singular (with blow-ups).
One can then define a graph whose vertices are the fixed points and where two fixed points are connected
by an edge if and only if the $\frak{sl}_2(\R)$ flow between them is regular.  
This turns out to be the incidence
graph that gives the integral cohomology of the real flag variety.
The total number of blow-ups in the Toda flows
is related to the polynomial associated with the rational cohomology of a
certain compact subgroup \cite{C:89, CK:06, CK:07}.

\section{Early versions of the finite nonperiodic real Toda lattice}\label{Sec:FiniteNonperiodicReal}

Consider $n$ particles, each with mass 1, arranged along a line at positions $q_1, ..., q_n$.
Between each pair of adjacent particles, there is a force whose magnitude depends exponentially on the distance 
between them.  Letting $p_k$ denote the momentum of the $k$th particle, and noting that $\frac{d}{dt}q_k = p_k$ since 
each mass is 1, the total energy of the system is the Hamiltonian
\begin{equation}
H = \frac{1}{2} \sum_{k = 1}^n\, p_k^2 + \sum_{k = 1}^{n-1} \,e^{-(q_{k+1} - q_k)} \ .
\label{Hamiltonian} \end{equation}
The equations of motion 
\begin{equation}\label{Hamiltonian DEs}
\displaystyle{\frac{dq_k}{dt} = \frac{\partial H}{\partial p_k} } \hspace{.5cm}  \mbox{and} \hspace{.5cm} \displaystyle{\frac{dp_k}{dt} = -\frac{\partial H}{\partial q_k}}
\end{equation}
give the system of equations for the finite nonperiodic Toda lattice,
\begin{equation}\label{Toda Equations}
\left\{\begin{array}{lllllll}
\displaystyle{\frac{dq_k}{dt}=p_k,} & k = 1, ..., n,  \\[2.0ex]
\displaystyle{\frac{dp_k}{dt} = -e^{-(q_{k+1} - q_k)} + e^{-(q_k - q_{k-1})},} \ \ \ & k = 1, ..., n.
\end{array}\right.
\end{equation}
Here we set $e^{-(q_1 - q_0)} = 0$ and $e^{-(q_{n+1} - q_n)} = 0$
with the formal boundary conditions
$q_0 = -\infty,$ and $q_{n+1} = \infty$.

\subsection{Symmetric form} \label{Symmetric Toda}

There are two classic Lax forms of Equations (\ref{Toda Equations}): the symmetric form and the Hessenberg form.
For the symmetric form, we make the change of variables (Flaschka \cite{Flaschka I}, Moser \cite{Moser})
\begin{equation}\label{Symmetric Variables}
\left\{\begin{array}{llll}
\displaystyle{a_k = \frac{1}{2}e^{-\frac{1}{2}(q_{k+1} - q_k)}}, \ \ \ & k = 1, ..., n-1\\[2.0ex]
\displaystyle{b_k = -\frac{1}{2}\,p_k,} \ \ \ & k = 1, ..., n \ .
\end{array}\right.
\end{equation}
In these variables, the Toda system (\ref{Toda Equations}) becomes
\begin{equation}\label{Symmetric Toda Equations}
\left\{ \begin{array}{llll}
\displaystyle{\frac{da_k}{dt} = a_k(b_{k+1} - b_k),} \ \ \ & k = 1, ..., n-1 \\[2.0ex]
\displaystyle{\frac{db_k}{dt} = 2(a_k^2 - a_{k-1}^2),} \ \ \  & k = 1, ..., n  
\end{array}\right.
\end{equation}
with boundary conditions 
$a_0 = 0$ and $ a_n = 0$.
Because the $a_k$ are real exponential functions, they are strictly positive for all time.

\begin{remark}\label{Remark:signs}
Making a change in sign $a_k\leftrightarrow -a_k$ for one or more values of $k$  in Definition \ref{Symmetric Variables} does not change Equations
\eqref{Symmetric Toda Equations}. That is, the systems $(\pm a_k,b_k)$ are equivalent for all choices of signs.
\end{remark}

The system \eqref{Symmetric Toda Equations} can be written in Lax form as 
\begin{equation}
\frac{d}{dt}L(t) = [\Pi_{\frak{so}}(L(t)),L(t)]
 \label{Symmetric Lax Equation}
\end{equation}
where $L$ is the symmetric tridiagonal matrix, and $\Pi_{\frak{so}}(L)$ is the skew-symmetric projection of $L$,
\begin{equation}
L = \left( \begin{array}{cccc} b_1 & a_1 &  & \\ 
a_1 & \ddots & \ddots &  \\
 & \ddots & \ddots & a_{n-1} \\
 &  & a_{n-1} & b_n
\end{array}  \right)\qquad\text{and}\qquad  \Pi_{\frak{so}}(L)=(L)_{>0}-(L)_{<0}.
\label{Symmetric}
\end{equation} 
Here $(L)_{>0}$ (resp. $(L)_{<0}$) is the strictly upper (resp. lower) triangular matrix of $L$.

Any equation in the Lax form $\frac{d}{dt}L = [B, L]$ for matrices $L$ and $B$ has the 
immediate consequence that the flow
preserves the spectrum of $L$.  To check this, it suffices to show that the function $\mbox{tr}(L^k)$, the trace of $L^k$,  is constant for each $k$.
One shows first by induction that $\frac{d}{dt}L^k= [B, L^k]$ and then observes that 
$\frac{d}{dt}[\mbox{tr} (L^k)]  =  \mbox{tr}[\frac{d}{dt}(L^k)] =  \mbox{tr}[B, L^k] = 0$.
We now have $n-1$ independent invariant functions
$$H_k(L) = \frac{1}{ k + 1}  \ \mbox{tr} L^{k + 1} \,.$$
The Hamiltonian (\ref{Hamiltonian}) is related to $H_1(L)$ by 
$H=4H_1(L)$
with the change of variables (\ref{Symmetric Variables}).

A property of real tridiagonal symmetric matrices (\ref{Symmetric}) with $a_k \neq 0$ for all $k$ is that the eigenvalues $\lambda_k$ are real and distinct.  Let $\Lambda$ be a set of $n$ real distinct eigenvalues, and let ${\mathcal M}_\Lambda = \{L \text{ in } \eqref{Symmetric} : \mbox{spec}(L) = \Lambda \}$.  Then
 ${\mathcal M}_{\Lambda}$ is a symplectic manifold.  Each invariant function $H_k(L)$ generates a Hamiltonian flow via the symplectic structure, and the flows are involutive with respect to that structure (see \cite{Arnold} for the general framework and \cite{Notes} for the Toda lattice specifically).  
In Section \ref{Full Symmetric real Toda lattice}, we describe the Lie-Poisson structure for the Equations (\ref{Symmetric Lax Equation}). 

Moser \cite{Moser} analyzes the dynamics of the Toda particles, showing that  for any initial configuration, $q_{k+1} - q_k$ tends to 
$\infty$ as $t \to \pm \infty$.  Thus,  
the off-diagonal entries of $L$ tend to zero as $t \to \pm \infty$ so that
$L$ tends to a diagonal matrix whose diagonal entries are the eigenvalues.
We will order them as $\lambda_1 < \lambda_2 < \cdots < \lambda_n$.
The analysis in \cite{Moser} shows the sorting property of the eigenvalues,
\begin{equation}\label{sorting}
L(t) ~\longrightarrow~\left\{\begin{array}{llll}
 \mbox{diag}(\lambda_n, \lambda_{n-1}, \cdots, \lambda_1)\qquad & \text{as}~ t\to\infty\,,\\[1.0ex]
\mbox{diag}(\lambda_1, \lambda_2, \cdots, \lambda_n)\qquad &\text{as}~t\to-\infty\,.
\end{array}\right.
\end{equation}
 The physical interpretation of this is
that as $t \to -\infty$, the particles $q_k$ approach the velocities $p_k(-\infty) = -2\lambda_k$, and as $t \to \infty$, 
the velocities are interchanged so that $p_k(\infty) = -2\lambda_{n-k+1}$.
Asymptotically, the trajectories behave as
\[
\left\{\begin{array}{lll}
q_k(t) & \approx & \lambda_k^{\pm} t + c_k^{\pm},  \\[1.0ex]
p_k(t) & \approx & \lambda_k^{\pm} \ ,
\end{array}\right.\hskip1.5cm \text{as}~ t\to\pm\infty\,.
\]
where $\lambda_k^+ = \lambda_k$ and $\lambda_k^- = \lambda_{n-k+1}$.  

Symes solves the Toda lattice using the QR-factorization;
his solution, which he verifies in \cite{Symes:82} and proves in a more general context in \cite{Symes:80},
is equivalent to the following.  To solve (\ref{Symmetric Toda Equations}) with initial matrix $L(0)$,
take the exponential $e^{tL(0)}$ and use Gram-Schmidt orthonormalization to factor it as
\begin{equation}
e^{tL(0)} = k(t)r(t) \ , \label{orthog x upper}
\end{equation}
where $k(t) \in SO(n)$ and $r(t)$ is upper triangular.
Then the solution of (\ref{Symmetric Toda Equations}) is
\begin{equation}\label{TodaL}
L(t) = k^{-1}(t) L(0) k(t)=r(t)L(0)r^{-1}(t) \ .
\end{equation}
Since the Gram-Schmidt orthonormalization of $e^{tL(0)}$ can be done for all $t$, this shows that
the solution of the Toda lattice equations (\ref{Symmetric Toda Equations}) on the set of
symmetric tridiagonal matrices $L$ of \eqref{Symmetric}
is defined for all $t$.  

We mention also the $\tau$-functions, which play a key role of the theory of
integrable systems (see for example \cite{H:04,MJD:00}). Let us first introduce the following symmetric
matrix, called the moment matrix,
\begin{equation}\label{Moment Toda}
M(t):=e^{2tL(0)}=r^T(t)k^T(t)k(t)r(t)=r^T(t)r(t)\,,
\end{equation}
where $r^T$ denotes the transpose of $r$, and note that $k^T=k^{-1}$.
The decomposition of a symmetric matrix to an upper-triangular matrix times its transpose on the left is called the Cholesky factorization.
This factorization is used to find the matrix $r$, and then the matrix $k$ can be found
by $k=e^{tL(0)}r^{-1}$.  The $\tau$-functions, $\tau_j$ for $j=1,\ldots, n-1$, are defined by
\begin{equation}\label{Todatau}
\tau_j(t):={\rm det}\,(M_j(t))=\prod_{i=1}^j r_i(t)^2\,, 
\end{equation}
where $M_j$ is the $j\times j$ upper-left submatrix of $M$, and we denote ${\rm diag}(r)={\rm diag}(r_1\ldots,r_n)$. 
We see from (\ref{TodaL}), i.e.
$L(t)r(t)=r(t)L(0)$, that
we have
\[
a_j(t)=a_j(0)\frac{r_{k+1}(t)}{r_{k}(t)}\,.
\]
Since $r_k(t)\ne 0$ for all $k$, the signs $a_j(t)$ remain the same.
With (\ref{Todatau}) and \eqref{Symmetric Toda Equations}, we obtain
\begin{equation}\label{a}
a_j(t)=a_j(0)\frac{\sqrt{\tau_{j+1}(t)\tau_{j-1}(t)}}{\tau_j(t)}
\qquad \text{and}\qquad b_j(t)=\frac{1}{2}\frac{d}{dt}\ln\left(\frac{\tau_j(t)}{\tau_{j-1}(t)}\right)\,.
\end{equation}
One should note that the $\tau$-functions are just defined from the moment matrix $M=e^{2tL(0)}$,
and the solutions $(a_j(t), b_j(t))$ are explicitly given by those $\tau$-functions {\it without}
the factorization.

\subsection{Hessenberg form}

The symmetric matrix $L$ in (\ref{Symmetric}), when conjugated by the diagonal matrix 
$D = \mbox{diag}(1, a_1, \ldots, a_{n-1})$, gives a matrix $Y = D L D^{-1}$ in Hessenberg form:
\begin{equation}
Y = \left( \begin{array}{cccc} b_1 & 1 &  & \\ 
a_1^2 & \ddots & \ddots &  \\
 & \ddots & \ddots & 1 \\
 &  & a_{n-1}^2 & b_n
\end{array} \right) \ . \label{Hessenberg Y}
\end{equation}
The Toda equations \eqref{Symmetric Lax Equation} now take the Lax form for $X:=2Y$,
\begin{equation}
\frac{d}{dt}X = [X, (X)_{<0}]=[(X)_{\ge 0},X], \label{Hessenberg equation}
\end{equation}
where $(X)_{\ge 0}=X-(X)_{<0}$ is the upper-triangular part of $X$.
Equation (\ref{Hessenberg equation}) with 
\begin{equation}
X = \left( \begin{array}{cccc} f_1 & 1 &  & \\ 
g_1 & \ddots & \ddots &  \\
 & \ddots & \ddots & 1 \\
 &  & g_{n-1} & f_n
\end{array} \right)  \label{Hessenberg}
\end{equation}
is called the Hessenberg form of the nonperiodic Toda lattice.
Again, since the equations are in Lax form, the functions 
$H_k(X) = \frac{1}{ k + 1}  \ \mbox{tr} X^{k + 1} \,$ are constant in $t$.  

Notice that the Hessenberg and symmetric Lax formulations of (\ref{Toda Equations})
are simply different
ways of expressing the same system.  The solutions exist for all time and exhibit the same behavior as $t \to \pm \infty$.  However, when we allow the subdiagonal entries to assume any real value, the
symmetric and Hessenberg forms differ in their geometry and topology and in the character of their solutions.  

\subsection{Isospectral manifolds in the real tridiagonal symmetric form}\label{SSec:ExtendedSymmTridiag}
Here we consider the Lax equation \eqref{Symmetric Lax Equation} where 
the $a_k$ in the symmetric Lax matrix
$L$ may be any real numbers.  As mentioned in Remark \ref{Remark:signs}, the equations with different signs in the $a_k$
are the same.  In particular, if $a_k\ne 0$ for all $k$, then the eigenvalues are real and distinct.

Let $\mathcal{M}_{\Lambda}$ denote the set of $n \times n$ matrices of the form (\ref{Symmetric}) with fixed eigenvalues 
$\lambda_1 < \lambda_2 < \cdots < \lambda_n$.  $\mathcal{M}_{\Lambda}$ contains $2^{n-1}$ components of dimension $n-1$, 
where each component consists of
all matrices in $\mathcal{M}_{\Lambda}$ with a fixed choice of sign for each $a_k$.   The solution of (\ref{Symmetric Lax Equation}) with initial condition in a given 
component remains in that component for all $t$, because the solutions preserve the sign of each $a_k$. 
Each lower-dimensional component, where one or more $a_k$ is zero and the signs of the other $a_k$ are fixed,
is also preserved by the Toda flow through any initial matrix in that component.
Tomei \cite{Tomei} shows that $\mathcal{M}_{\Lambda}$ is a compact smooth manifold of dimension $n-1$ 
containing $2^{n-1}$ open components, each diffeomorphic to ${\mathbb{R}}^{n-1}$ (see also \cite{Tomei2}).  
On each component, $a_k \neq 0$ for all $k$, and the sign of each $a_k$ is fixed.   The components are glued together along the 
lower-dimensional sets where one or more $a_k$ is zero. 

\begin{figure}[t!]
\centering
\includegraphics[scale=0.9]{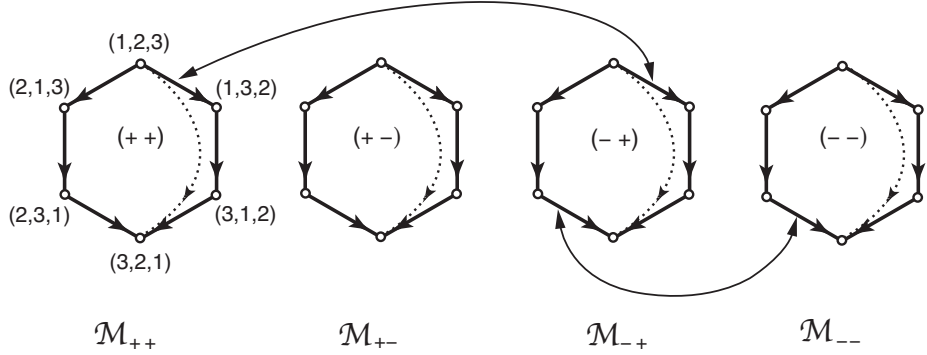}
\caption{The Tomei manifold $\mathcal{M}_{\Lambda}$ for the symmetric tridiagonal $\frak{sl}_3(\Real)$ Toda lattice.
The 3-tuples $(i,j,k)$ on the vertices indicate the diagonal matrices $L=\mbox{diag}(\lambda_i,\lambda_j,\lambda_k)$.
Each hexagon $\mathcal{M}_{\epsilon_1,\epsilon_2}$ corresponds to the moment polytope
(see Section \ref{moment}) for the Toda lattice with the signs $(\epsilon_1,\epsilon_2)=({\rm sgn}(a_1),{\rm sgn}(a_2))$. 
The boundaries correspond to the $sl(2,\Real)$ Toda lattices associated with either $a_1=0$ or $a_2=0$.
$\mathcal{M}_{\Lambda}$ results from gluing corresponding edges of the hexagons. For example,
the edge between $(1,2,3)$ and $(1,3,2)$ in $\mathcal{M}_{++}$ is glued with the same edge in
$\mathcal{M}_{-+}$, since this edge indicates $a_1=0$ and $a_2 > 0$. 
In the other gluing shown, $a_2=0$ and $a_1 < 0$. }
\label{fig:tomei}
\end{figure}

 For $n = 3$, there are four 2-dimensional
components, denoted as $\mathcal{M}_{++}, \mathcal{M}_{+-}, M_{-+}$, and $\mathcal{M}_{--}$, according to the signs of $a_1$ and $a_2$. 
The closure of each component is obtained by adjoining the six diagonal matrices 
${\rm diag}(\lambda_i,\lambda_j,\lambda_k)$ where all the $a_k$ vanish (these are the fixed points of the Toda flows)
and six 1-dimensional sets where exactly one $a_k$ is zero.  
The four principal components are glued together along the loci of $\mathcal{M}_{\Lambda}$ where one or more
$a_k$ vanish.   
In Figure \ref{fig:tomei}, we illustrate the Tomei manifold 
$\mathcal{M}_{\Lambda}$ for the $\frak{sl}_3(\Real)$ symmetric Toda lattice,
\[
\mathcal{M}_{\Lambda}=\overline{\mathcal{M}}_{++}\cup\overline{\mathcal{M}}_{+-}\cup\overline{\mathcal{M}}_{-+}\cup\overline{\mathcal{M}}_{--}\,,
\]
where the cups include the specific gluing according to the signs of the $a_k$.
The resulting manifold $\mathcal{M}_{\Lambda}$ is a connected sum of two tori, the compact Riemann surface of genus two. This can be easily seen from Figure \ref{fig:tomei} as follows:
Gluing those four hexagons, $\mathcal{M}_{\Lambda}$ consists of 6 vertices, 12 edges and 4 faces. Hence the Euler characteristic
is given by $\chi(\mathcal{M}_{\Lambda})=6-12+4=-2$, which implies that the manifold has genus $g=2$ (recall $\chi=2-2g$).
The fact that $\mathcal{M}_{\Lambda}$ is orientable can be shown by giving
an orientation for each hexagon so that the directions of two edges in the gluing cancel each other.
Since compact two-dimensional surfaces are completely characterized by
their orientability and Euler characters, we conclude that
the manifold $\mathcal{M}_{\Lambda}$ is a connected sum of two tori.

The Euler characteristic of $\mathcal{M}_{\Lambda}$ (for general $n$) is determined in  \cite{Tomei} as follows.  Let 
$L = \mbox{diag}(\lambda_{\sigma(1)}, ..., \lambda_{\sigma(n)})$ be a diagonal matrix in $\mathcal{M}_{\Lambda}$, where $\sigma$ is
a permutation of the numbers $\{1, ..., n\}$, and let $r(L)$ be the number of times that $\sigma(k)$ is less
than $\sigma(k+1)$.  Denote by $E(n,k)$ the number of diagonal matrices in $\mathcal{M}_{\Lambda}$ with $r(L) = k$.
Then the Euler characteristic of $\mathcal{M}_{\Lambda}$ is the alternating sum of the $E(n,k)$:
$$\chi(\mathcal{M}_{\Lambda}) = \sum_{k=0}^{n} (-1)^k E(n,k) \ .$$

An isospectral set where the eigenvalues are not distinct is not a manifold.  
For example, the isospectral set with spectrum $(1,1,3)$ has the 
shape of a figure eight  \cite{Tomei, CK:04}.

\subsection{Indefinite Toda lattice in real tridiagonal Hessenberg form}\label{Extended Hessenberg Toda}

We return to the Hessenberg form with $X$ as in 
(\ref{Hessenberg}), and allow the $g_k$ to assume arbitrary real values.
Recall that in the formulation of the original Toda equations, 
all the $g_k$ were positive, so that the eigenvalues were real and distinct.  When $g_k \neq 0$ for some $k$, the eigenvalues may be complex or may coincide.  Even in the case where all
the eigenvalues are real and distinct, the case with some $g_k<0$ causes blow-ups in the flows so that the topology of the isospectral manifolds is very different from the topology of the Tomei manifolds described in the previous section \cite{KY:98,FT:97}.

The matrices of the form (\ref{Hessenberg}) with $g_k \neq 0$ for all $k$ are partitioned into $2^{n-1}$ different Hamiltonian systems,
each determined by a choice of signs of the $g_k$.   Letting $\sigma_k = \pm 1$ for $k = 1, ..., n$
and taking the sign of $g_k$ to be $\sigma_k \sigma_{k+1}$, Kodama and Ye \cite{KY:96} give the Hamiltonian
for the system with this choice of signs as 
\begin{equation}
H = \frac{1}{2} \sum_{k = 1}^n\, y_k^2 + \sum_{k = 1}^{n-1} \sigma_k \sigma_{k+1} e^{-(x_{k+1} - x_k)} \ ,
\label{Indefinite Hamiltonian} \end{equation}
where the variables $(f_k,g_k)$ in the Hessenberg form are given by
\begin{equation}\label{indefinite fg}
\left\{\begin{array}{llllll}
\displaystyle{f_k = -\frac{1}{2}\, y_k }\ , & k = 1, ..., n \\[1.5ex]
\displaystyle{g_k = \frac{1}{4}\,\sigma_k \sigma_{k+1}\, e^{-(x_{k+1} - x_k)},} \ \ \ & k = 1, ..., n-1 \ . 
\end{array}\right.
\end{equation}
The system (\ref{Hessenberg equation}) with the Hamiltonian $H$ in \eqref{Indefinite Hamiltonian} is called the {\it  indefinite} Toda lattice.
The negative signs in (\ref{Indefinite Hamiltonian}) correspond to attractive forces between adjacent particles,
which causes the system to become undefined at finite values of $t$, as is seen in the solutions obtained in \cite{KY:96} and \cite{KY:98} by inverse scattering.

The blow-ups in the solutions are also apparent in the factorization solution of the Hessenberg form.
To solve (\ref{Hessenberg equation}) with initial condition $X(0)$, we consider the LU-factorization
of the exponential $e^{tX(0)}$,
\begin{equation}
e^{tX(0)} = n(t)b(t) \ , \label{lower x upper}
\end{equation}
where $n(t)$ is lower unipotent and $b(t)$ is upper-triangular.
Then, as shown in \cite{RSTS, Reiman} (see also \cite{Guest,Perelomov}), 
\begin{equation}
X(t) = n^{-1}(t) X(0) n(t)=b(t) X(0) b^{-1}(t) \label{factorization solution}
\end{equation}
solves (\ref{Hessenberg equation}).
Notice that the factorization (\ref{lower x upper}) is obtained by Gaussian elimination, which multiplies 
$e^{tX(0)}$ on the left by elementary row operations to put it in upper-triangular form.  This process works only
when all principal minors (the determinants of upper left $k \times k$ blocks, which are the 
$\tau$-functions as defined in \eqref{Todatau}) are nonzero.
At particular values of $t \in {\mathbb{R}}$, this factorization can fail, and the solution (\ref{factorization solution})
becomes undefined.

The solutions $(f_k,g_k)$ can be expressed in terms of the $\tau$-functions 
\begin{equation}\label{indefinite tauk}
\tau_k(t):=  \left[e^{tX(0)}\right]_k=\prod_{j=1}^k d_{j}(t)\,,
\end{equation}
where $[e^{tX(0)}]_k$ is the $k\times k$ principal minor of $e^{tX(0)}$, and
${\rm diag}(b)={\rm diag}(d_1,\ldots,d_n)$. As in the previous case of symmetric Toda, from (\ref{factorization solution}), we have
\begin{equation}\label{g indefinite}
g_k(t)=g_k(0)\frac{\tau_{k+1}(t)\tau_{k-1}(t)}{\tau_k(t)^2}\qquad\text{and}\qquad 
f_k(t)=\frac{d}{dt}\ln\left(\frac{\tau_k(t)}{\tau_{k-1}(t)}\right)\,.
\end{equation}
Now it it clear that the factorization (\ref{lower x upper}) fails if and only if $\tau_k(t)=0$ for some $k$.
Then a blow-up (singularity) of the system (\ref{Hessenberg equation}) can be characterized by
the zero sets of the $\tau$-functions.

\begin{example}\label{example 1.4.1} To see how blow-ups occur in the factorization solution, consider the initial matrix
$$X_0 =  \left( \begin{array}{rr} 1 & 1 \\ 
 -1 & -1 \end{array} \right) \ .$$
When $t \neq -1$,
$$e^{t X_0} = \left( \begin{array}{cc} 1+t & t \\ 
 -t & 1-t \end{array} \right) 
= \left( \begin{array}{cc} 1 & 0 \\ 
 \frac{-t}{1 + t} & 1 \end{array} \right) \ 
\left( \begin{array}{cc} 1+t & t \\ 
 0 & \frac{1}{1+t} \end{array} \right) \ , $$
and the solution evolves as in (\ref{factorization solution}).
The $\tau$-function is given by $\tau_1(t)=1+t$, and
when $t = -1$, this factorization does not work.  However,
we can multiply $e^{- X_0}$ on the left by a lower unipotent matrix $n^{-1}$ (in this case the identity)
to put it in
the form $w b$, where $w$ is a permutation matrix:
$$e^{- X_0} = \left( \begin{array}{cc} 0 & -1 \\ 
 1 & 2 \end{array} \right) \ = \ 
\left( \begin{array}{cc} 1 & 0 \\ 
 0 & 1 \end{array} \right) 
\left( \begin{array}{cc} 0 & -1 \\ 
 1 & 0 \end{array} \right) \ 
\left( \begin{array}{cc} 1 & 2 \\ 
 0 & 1 \end{array} \right)  \ .$$
This example will be taken up again in Section \ref{Complex tridiagonal Hessenberg}, where it is shown how the factorization using a permutation matrix
leads to a compactification of the flows.  
\end{example}

In general, when the factorization (\ref{lower x upper}) is not possible at time $t = \bar{t}$, 
$e^{\bar{t}X(0)}$ can be factored as
$e^{\bar{t}X(0)} = n(\bar{t}) \ w \ b(\bar{t})$ for some permutation matrix $w$.  
In \cite{EFH}, this factorization is used to complete the flows 
(\ref{factorization solution}) 
through the blow-up times by embedding them into a flag variety.  
We examine this further in the context of the
complex tridiagonal Hessenberg form.

To describe the topology of a generic isospectral set $\mathcal{M}_{\Lambda}$ 
in this version of the Toda lattice, 
it is first shown that because of the blow-ups in $X$,
$\mathcal{M}_{\Lambda}$ is a noncompact manifold of dimension $n-1$ \cite{KY:98}.
The manifold is compactified by completing the flows through the blow-up times.  The $2 \times 2$ case is basic to the compactification for general $n$.
The set of $2\times 2$ matrices with fixed eigenvalues $\lambda_1 < \lambda_2$,
\begin{equation}
\mathcal{M}_{\Lambda} = \Bigg\{ \left( \begin{array}{cc} f_1 & 1 \\ 
 g_1 & f_2 \end{array} \right) :  \lambda_1 < \lambda_2 \  \Bigg\} \ , \label{2x2}
\end{equation}
consists of two components, $\mathcal{M}_{+}$ with $g_1 > 0$ and $\mathcal{M}_{-}$ with $g_1 < 0$, together with two
fixed points, 
$$X_1 = \left( \begin{array}{cc} \lambda_1 & 1 \\ 
 0 & \lambda_2 \end{array} \right) \hspace{.4cm} \mbox{and} \ \ \ 
X_2 = \left( \begin{array}{cc} \lambda_2 & 1 \\ 
 0 & \lambda_1 \end{array} \right) \ .$$
Writing $f_2 = \lambda_1 + \lambda_2 - f_1$ and substituting this into the equation for the determinant,
$f_1 f_2 - g_1 = \lambda_1 \lambda_2$, shows that $M_{\lambda}$ is the parabola
\begin{equation}
g_1 = -(f_1 - \lambda_1)(f_1 - \lambda_2) \label{parabola} \ .
\end{equation}
This parabola opens down, crossing the axis $g_1 = 0$ at $f_1 = \lambda_1$ and $f_1 = \lambda_2$, corresponding
to the fixed points $X_1$ and $X_2$.
For an initial condition with $g_1 > 0$, the solution is defined for all $t$; it flows away from $p_2$ toward
$p_1$.  This illustrates what is known as the sorting property, which says that as $t \to \infty$, 
the flow tends toward the fixed point with the eigenvalues in decreasing order along the diagonal.
The component with $g_1 < 0$ is separated into disjoint parts, one with $f_1 < \lambda_1$ and the other
with $f_1 > \lambda_2$.   The solution starting at an initial matrix with $f_1 > \lambda_2$ flows toward the fixed
point $X_2$ as $t \to \infty$.  For an initial matrix with $f_1 < \lambda_1$, the solution flows away from
$X_1$, blowing up at a finite value of $t$.  By adjoining a point at infinity to connect these two branches of the 
parabola, the flow is completed through the blow-up time and the resulting manifold is the circle, $S^1$.

\begin{figure}[t!]
\centering
\includegraphics[scale=0.9]{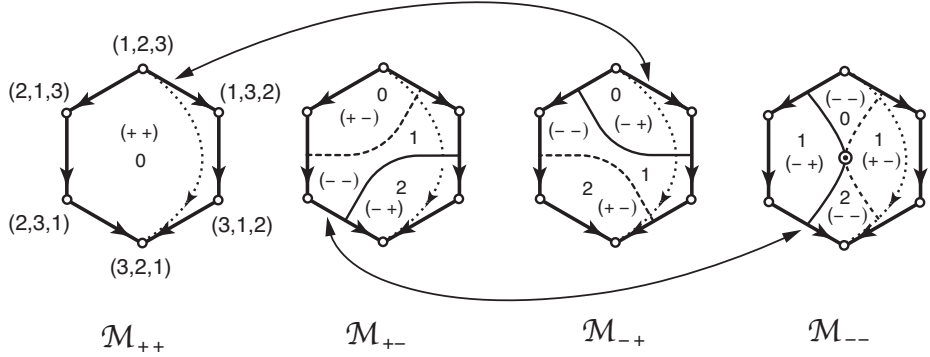}
\caption{The compactification of the isospectral manifold $\mathcal{M}_{\Lambda}$
for the indefinite $\frak{sl}_3(\Real)$ Toda lattice. Each hexagon
indicates the moment polytope associated with an indefinite Toda lattice.
The signs $(\epsilon_1,\epsilon_2)$ in $\mathcal{M}_{\epsilon_1,\epsilon_2}$ are those of $(g_1,g_2)$ as $t\to-\infty$, and the signs in the hexagons indicate the signs of $(g_1,g_2)$.
The gluing rule according to the sign changes of $g_i$ is the same as that in the Tomei manifold, but the pattern is now different.
For example, the edge between $(2,3,1)$ and $(3,2,1)$ in $\mathcal{M}_{--}$ is now glued with
that in $\mathcal{M}_{+-}$. The solid and dashed lines in the hexagons
show the points where the solutions blow up; $\tau_1=0$ (solid) and $\tau_2=0$ (dashed). 
The numbers in the sections indicate the number of blow-ups along the flow from
$t=-\infty$ to $+\infty$ (see Section \ref{blow-ups}).}
\label{fig:A2indefinite}
\end{figure}

For general $n$, the manifold $\mathcal{M}_{\Lambda}$ with spectrum $\Lambda$
contains $n!$ fixed points of the flow, where the eigenvalues are arranged along the diagonal.  
These vertices are connected to each other by incoming and outgoing edges analogous to the flows connecting the two
vertices when $n = 2$.  
The result is nonorientable for $n > 2$.  For $n = 3$, it is a connected sum of two Klein bottles. 
Figure \ref{fig:A2indefinite} illustrates the compactification of $\mathcal{M}_{\Lambda}$ for
the $\frak{sl}_3(\Real)$ indefinite Toda lattice. With this gluing, the compactified manifold $\overline{\mathcal{M}}_{\Lambda}$ has Euler characteristic $\chi(\overline{\mathcal{M}}_{\Lambda})=-2$
as in the case of the Tomei manifold (see Figure \ref{fig:tomei}). The non-orientability is seen in the
non-cancellation of the given orientations of the hexagons.

Casian and Kodama \cite{CK:02} (see also \cite{CK:02b}) show that the compactified isospectral manifold is identified as a connected completion of the disconnected Cartan subgroup of $G=Ad(SL(n,\mathbb{R})^{\pm})$.  It is diffeomorphic to a toric variety in the flag variety of $G$.
We give more details in Section \ref{blow-ups}.


\subsection{Full symmetric real Toda lattice}  \label{Full Symmetric real Toda lattice}

In 1986, the paper \cite{DLNT:86} by Deift, Li, Nanda, and Tomei brought a radical departure from the tridiagonal Toda lattices that had been heretofore studied, by expanding the phase space to the set of full symmetric matrices.
Consider the symmetric Toda equation
\begin{equation}
\frac{d}{dt}L = [\, \Pi_{\frak{so}}(L), \ L] \qquad\text{with}\qquad \Pi_{\frak{so}}(L)=(L)_{>0}-(L)_{<0}\, . \label{full symmetric Toda}
\end{equation}
as in (\ref{Symmetric Lax Equation}), where $L$ is now a full symmetric matrix with distinct eigenvalues.  
The authors of \cite{DLNT:86} show that (\ref{full symmetric Toda}) remains completely integrable for the generic case.
They present a sufficient number of constants of motion in involution and construct the associated angle variables.
The additional constants of motion are found by a chopping construction on the matrix that was later extended in \cite{EFS} to the complex full Kostant-Toda lattice, which we describe in more detail in Section \ref{full Kostant-Toda}.

The Lie-Poisson structure on the space of symmetric real matrices is the Kostant-Kirillov form (as explained in \cite{DLNT:86}), with respect to which the 
Toda flows 
(\ref{full symmetric Toda}) may be expressed in Hamiltonian form as
\[
\frac{d}{dt}L=\{H_1,L\}(L) \qquad {\rm with}\qquad H_1(L)=\frac{1}{2}{\rm tr}(L^2)\,.
\]

Using the Poisson structure, we may extend (\ref{full symmetric Toda}) to define the Toda lattice hierarchy 
generated by the Hamiltonians $H_k(L)=\frac{1}{k+1}{\rm tr}(L^{k+1})$:
\begin{equation}
\frac{\partial}{\partial t_k}L=\{H_k,L\}(L)=[\Pi_{\frak{so}}\nabla H_k,L]\qquad\text{for}\quad k=1,2,\ldots,n-1. \label{Symmetric Toda Hierarchy}
\end{equation}
where $\text{tr}( X\nabla f)=\lim_{\epsilon\to 0}\frac{d}{d\epsilon}f(L+\epsilon X)$ so that $\nabla H_k=L^k$.
Each flow stays on a co-adjoint orbit in the phase space of the symmetric Toda, which is  Sym$(n):=\{L\in\frak{sl}_n(\mathbb{R}):L^T=L\}$.
The Poisson structure is nondegenerate when restricted to a co-adjoint orbit,
and the level sets of the integrals found in \cite{DLNT:86}
are the generic co-adjoint orbits.

In \cite{KM:96}, Kodama and McLaughlin give the explicit solution of the Toda lattice hierarchy (\ref{Symmetric Toda Hierarchy}) 
on full symmetric matrices with distinct eigenvalues by solving the inverse scattering problem of the system
\begin{align*}
L \,\Phi  =  \Phi\, \Lambda \qquad \text{and}\qquad
\frac{\partial}{\partial t_k}\Phi  =  \Pi_{\frak{so}}(L^k)\, \Phi 
\end{align*}
with $\Lambda={\rm diag}(\lambda_1,\ldots,\lambda_n)$.
Since $L$ is symmetric, the matrix $\Phi$ of eigenvectors is taken to be orthogonal:
$$
L = \Phi\, \Lambda\, \Phi^T
$$
with $\Phi = [\phi(\lambda_1), ..., \phi(\lambda_n)]$, where the $\phi(\lambda_k)$ is the normalized eigenvector of $L$ with eigenvalue $\lambda_k$.

The indefinite extension of the full symmetric Toda lattice  is studied in 
\cite{KM:96}, where explicit solutions of $\phi(\lambda_k,t)$ are obtained by inverse scattering.
The authors also give an alternative derivation of the solution using the factorization method of Symes \cite{Symes:82}, where
$e^{tL(0)}$ is factored into a product of a pseudo-orthogonal matrix times an upper triangular matrix (the HR-factorization).

\section{Complex Toda lattices}\label{Sec:ComplexTodaLattices}

 The jump from real to complex Toda lattices brought with it a powerful new tool for understanding the geometry of the iso-spectral varieties, namely, embeddings of the isospectral sets into the
 flag varieties. Under these mappings, the Toda flows generate group actions and blow-ups are compactified.
The geometry and topology of the compactified isospectral sets can then be described in terms of the moment map and moment polytope of the flag variety.  

\subsection{The moment map}\label{moment}
Let $G$ be a complex
semisimple Lie group, $H$ a Cartan subgroup of $G$, and $B$ a Borel
subgroup containing $H$.  If $P$ is a parabolic subgroup of $G$ that
contains $B$, then $G/P$ can be realized as the orbit of $G$ through
the projectivized highest weight vector in the projectivization, ${\mathbb P}(V)$, of an irreducible
 representation $V$ of $G$. 
Let ${\mathcal A}$ be the set of weights of $V$, counted with multiplicity;
the weights belong to $\frak{h}^*_{\mathbb{R}}$, the real part of the dual of the Lie algebra $\frak{h}$ of $H$.
Let $\{v_{\alpha}:\alpha \in {\mathcal A}\}$ be a basis of $V$ consisting of weight vectors.
A point [X] in $G/P$, represented by $X \in V$, has homogeneous
coordinates $\pi_{\alpha}(X)$, 
where $X = \sum_{\alpha \in {\mathcal A}} \pi_{\alpha}(X) v_{\alpha}$.
The moment map as defined in \cite{Kirwan} sends $G/P$ into $\frak{h}^*_{\mathbb{R}}$:
\begin{equation}\label{moment map}\begin{array}{cccccc}
\mu &: &G/P & \longrightarrow & \frak{h}^*_{\mathbb{R}}  \\[1.5ex]
& & \mbox{}[X] & \longmapsto & 
\displaystyle{\frac{\sum_{\alpha \in {\mathcal A}} |\pi_{\alpha}(X)|^2 \alpha}{\sum_{\alpha \in {\mathcal A}} |\pi_{\alpha}(X)|^2} } 
\end{array}\end{equation}
Its image is the weight polytope of $V$, also referred to as
the moment polytope of $G/P$.   

The fixed points of $H$ in $G/P$ are the points in the orbit of
the Weyl group $W$ through the projectivized highest weight vector
of $V$;
they correspond to the vertices of the polytope under the moment map.
Let $\overline{H \cdot [X]}$ be the closure of the orbit of $H$ 
through $[X]$.  Its image under $\mu$ is the convex hull of the 
vertices corresponding to the fixed points contained in $\overline{H \cdot [X]}$; 
these vertices are the weights $\{\alpha \in W \cdot \alpha^{V} :
\pi_{\alpha}(X) \neq 0 \}$, where $\alpha^{V}$ is the highest weight
of $V$ \cite{A:82}.
In particular, the image of a generic orbit, where no $\pi_{\alpha}$ vanishes, is the full polytope.
The real dimension of the image is equal to the complex dimension of
the orbit.


For $G = SL(n,{\mathbb C})$, $B$ the upper triangular
subgroup, and $H$ the diagonal torus.  The choice of $B$ determines a 
splitting of the root system into positive and negative roots and a system
$\Delta$ of simple roots.  The simple roots are ${\sf L}_i - {\sf L}_{i+1}$, where $i = 1, ..., n-1$ and ${\sf L}_i$ is a weight of the standard representation of $\frak{sl}_n$, i.e. for $h=\text{diag}(h_1,\ldots,h_n)\in\frak{h}$, ${\sf L}(h)=h_i$.  Then let $\frak{h}_\R^*$ denote the dual of $\frak{h}$,
\begin{equation}\label{dualCartan}
\mathfrak{h}_{\R}^*:=\text{Span}_{\R}\left\{\mathsf{L}_1,\ldots, \mathsf{L}_n~\Big|~\sum_{j=1}^n\mathsf{L}_j=0\right\}\cong\R^{n-1}.
\end{equation}
The Weyl group $W=\Sym_n$ acts by permuting the weight ${\sf L}_i$, and the moment polytope
of $G/B$ is  the convex hull of the weights ${\sf L}_{i_1,\ldots,i_n}$
for $(i_1,\ldots,i_n)=\pi(1,\ldots,n)$ with $\pi\in\Sym_n$, which is given by
\begin{equation}\label{weights}
\mathsf{L}_{i_1,\ldots,i_n}:=(n-1)\mathsf{L}_{i_1}+(n-2)\mathsf{L}_{i_2}+\cdots+\mathsf{L}_{i_{n-1}}.
\end{equation}
where the highest weight is ${\sf L}_{1,2,\ldots,n}$.
This moment polytope is referred to as the permutohedron, which is given by
\begin{equation}\label{perm}
\Perm_n=\CH\{\mathsf{L}_{\pi(1,\ldots,n)}\in\mathfrak{h}_{\R}^*~|~\pi\in \Sym_n\}.
\end{equation}


\subsection{Complex tridiagonal Hessenberg form} \label{Complex tridiagonal Hessenberg}
Let ${\mathcal M}$ be the set of complex tridiagonal Hessenberg matrices
of the form (\ref{Hessenberg}),
where the $f_k$ and $g_k$ are arbitrary complex numbers.
As before, the Toda flow is defined by (\ref{Hessenberg equation}) and the eigenvalues (equivalently, the traces of the powers of $X$)
are constants of motion.  The Hamiltonian $H_k(X) = \frac{1}{k + 1}  \ \mbox{tr} \left(X^{k + 1}\right)$
generates the flow
\begin{equation}
\frac{\partial X}{\partial t_k} =[X,\ (\nabla H_k)_{<0}]=[X,  \ (X^k)_{<0}] \ .  \label{Hessenberg hierarchy}
\end{equation}
The solution of (\ref{Hessenberg hierarchy}) can be found by the LU-factorization as in (\ref{lower x upper}).  That is, with 
$e^{t_kX^k(0)} = n(t_k)b(t_k)$, we have
Then \begin{equation}
X(t_k) = n^{-1}(t_k) X(0) n(t_k)=b(t_k)X(0)b^{-1}(t_k). \label{general factorization solution}
\end{equation}

Fix the eigenvalues $\lambda_j$, and consider the level set 
${\mathcal M}_{\Lambda}$ consisting of all matrices in ${\mathcal M}$ with spectrum $\Lambda=\{\lambda_1, ..., \lambda_n\}$ (we often identify $\Lambda=\text{diag}(\lambda_1,\ldots,\lambda)$.
In contrast to the real tridiagonal flows described in Section \ref{Extended Hessenberg Toda}, when $X$ is complex, ${\mathcal M}_{\Lambda}$ is no longer partitioned by signs of the $g_k$.  There is only one maximal component
where no $g_k$ vanishes.    The $n-1$ flows through any
initial $X$ with $g_k \neq 0$ for all $k$ generates the whole component.

In the case of distinct eigenvalues, Ercolani, Flaschka and Haine in \cite{EFH} construct a minimal nonsingular compactification of ${\mathcal M}_{\Lambda}$ on which the
flows (\ref{Hessenberg hierarchy}) extend to global holomorphic flows.  
The compactification is induced by an embedding of ${\mathcal M}_{\Lambda}$ into the flag variety $SL(n,{\mathbb C})/B$ with $B$,
the set of weakly upper-triangular matrices.  
The embedding depends on the following factorization by Kostant \cite{Kostant} of $X \in {\mathcal M}_{\Lambda}$.
Let $\epsilon_{\Lambda}$ be the matrix with $(\lambda_1, \ldots, \lambda_n)$ on the diagonal, 1's on the superdiagonal, and 0's elsewhere.  
Then 
every $X \in {\mathcal M}_{\Lambda}$ can be conjugated to $\epsilon_{\Lambda}$ by a unique element $n\in N$, the set of lower-triangular unipotent matrices:
\begin{equation}\label{Kostantfactorization}
X = n \epsilon_{\Lambda} n^{-1} \ .
\end{equation}

This defines a map of ${\mathcal M}_{\Lambda}$ into $G/B$:
\begin{equation}\label{embedding}\begin{array}{cccc}
j_{\Lambda} : & {\mathcal M}_{\Lambda} &\rightarrow &G/B\\[0.4ex]
&X & \mapsto & n^{-1} \ \mbox{mod} \ B \ . 
\end{array}\end{equation}
This mapping is an embedding \cite{Kostant Whittaker}, and the closure,
$\overline{j_{\Lambda}({\mathcal M}_{\Lambda})}$, 
of its image is a nonsingular and minimal compactification
of ${\mathcal M}_{\Lambda}$.   Let $n_0$ be the unique lower unipotent matrix such that $X(0) = n_0 \epsilon_{\Lambda} n_0^{-1}$.  Then
the solution (\ref{general factorization solution}) is 
$X(t_k) = n^{-1}(t_k) n_0 \epsilon_{\Lambda} n_0^{-1} n(t_k)$, where
$n_0^{-1} n(t_k)$ is lower unipotent.
The Toda flow $X(t_k)$ is mapped into the flag variety as
\begin{equation}\label{epsilon embedding}\begin{array}{cccc}
j_{\Lambda}(X(t_k)) & = & u_0^{-1} n(t_k) \ \mbox{mod} \ B \  \\ 
& = & u_0^{-1} e^{t_k X^k(0)} \ \mbox{mod} \ B \ . 
\end{array}\end{equation}
Even at values of $t_k$ where the first expression in (\ref{epsilon embedding}) is not defined because the LU-factorization of $e^{t_kX^k(0)}$ is not
possible, the second expression in (\ref{epsilon embedding}) is defined.  In this way, the embedding of $X(t_k)$ into $G/B$ completes the flows
through the blow-up times.  
This used in
\cite{EFH} to study the nature of the blow-ups of $X(t_k)$.
  
To illustrate this in a simple case, consider Example \ref{example 1.4.1} from Section \ref{Extended Hessenberg Toda}.
The isospectral set of $2 \times 2$ Hessenberg matrices with both eigenvalues zero is embedded 
into the flag variety $SL(2,{\mathbb C})/B$, which has the cell decomposition
\begin{equation}
{SL(2,{\mathbb C})}/{B} = \ {N B}/{B} 
\ \sqcup \
{N\left( \begin{array}{cc} 0 & -1 \\ 
 1 & 0 \end{array} \right) B}/{B} \ . \label{2x2 Bruhat}
\end{equation}
The big cell, ${NB}/{B}$, contains the image of the flow $X(t)$ whenever this flow is defined, that is, whenever
the factorization $e^{t X_0} = n(t) b(t)$ is possible.  
At $t = -1$, where $X(t)$ is undefined, the embedding $j_{\Lambda}$ completes the flow through the singularity.
The image $j_{\Lambda}(X(t))$ passes through
the flag $u_0^{-1} e^{- X(0)}$ at time $t = -1$, which is the cell on the right in (\ref{2x2 Bruhat}).

The cell decomposition (\ref{2x2 Bruhat}) is a special case of the cell stratification of $G/B$  known as the  
Bruhat decomposition.  This decomposition is defined in terms of the Weyl group $W$ as
\begin{equation}
G/B = \bigsqcup_{w \in W} N \dot{w} B / B\,, \label{Bruhat}
\end{equation}
where each component $N\dot{w}B/B$ is called the Bruhat cell associated to $w\in W$.
Here $\dot{w}$ is the representation of $w\in W$ on $G$.
For $G = SL(n, {\mathbb C})$, $W$ is the symmetric group of permutations $\Sym_n$, and 
$\dot{w}$ is the permutation matrix corresponding to $w\in\Sym_n$.
Thus, the Bruhat decomposition partitions flags according to which permutation matrix $\dot{w}$ is needed to perform the factorization
$g= n \dot{w} b$ for $g \in G$ with $n \in N$ and $b \in B$.  At all values of $t_k$ for which the flow $X(t_k)$ is defined, 
$j_{\Lambda}$ sends $X(t_k)$ into the big cell of the Bruhat
decomposition, since $\dot{w}$ is the identity matrix.  When the factorization $e^{t_kX^k(0)}=n(t_k)b(t_k)$ is not possible at time
$t_k = \bar{t}$,  the group element
$g=e^{\bar{t} X^{k}(0)}$ can be factored as
\begin{equation}
e^{\bar{t} X^{k}(0)} = n(\bar{t}) \dot{w} b(\bar{t}) \ , \label{w factorization}
\end{equation}
for some permutation matrix $\dot{w}$.  In this case, the flow (\ref{epsilon embedding}) enters the Bruhat cell $N \dot{w} B / B$ at time $t_k = \bar{t}$.  Ercolani et al.~in \cite{EFH} characterize the Laurent expansion of each pole of $X(t_1)$ in terms of the Bruhat cell that the solution enters
at the blow-up time.

The compactification of ${\mathcal M}_{\Lambda}$ where eigenvalues need not be distinct is studied in \cite{Complex Tridiagonal} by a modification of the embedding (\ref{embedding}) where the Hessenberg matrix $X$ is conjugated to a Jordan matrix $J$ of the spectrum (see \cite{Nongeneric}).
Since the Jordan matrix of $X$ has one block for each eigenvalue, all elements of ${\mathcal M}_{\Lambda}$ are conjugate.
Under the Jordan embedding, the maximal torus generated by the flows is diagonal if the eigenvalues are distinct and a product of a diagonal torus and a unipotent group
when eigenvalues coincide.  
Via the Jordan embedding, the $n-1$ flows $X(t_k) = n^{-1}(t_k) X(0) n(t_k)$ in (\ref{general factorization solution}),
with the appropriate choice of $X(0)$,
generate an action of the centralizer of $J$ in
$G = SL(n,{\mathbb C})$ on the flag variety $G/B$.  As shown in \cite{Nongeneric},
this group is a semi-direct product of the diagonal torus obtained
by setting all the entries above the diagonal equal to zero, and the unipotent group
obtained by setting all the diagonal entries equal to 1.
This group action, together with the moment map of the maximal torus, is used in \cite{Complex Tridiagonal} 
to identify each component in the boundary of the image of ${\mathcal M}_{\Lambda}$ in $G/B$
with a face of the moment polytope.

\subsection{The full Kostant-Toda lattice} \label{full Kostant-Toda}

In this version of the Toda lattice, Ercolani, Flaschka, and Singer \cite{EFS} connect the expanded phase space with its additional constants of motion introduced by Deift, Li, Nanda and Tomei in
\cite{DLNT:86} (see Section \ref{Full Symmetric real Toda lattice}) and the geometrically enlightening idea of embeddings of the isospectral sets into flag varieties introduced by Ercolani, Flaschka and Haine in 
\cite{EFH} (see Section \ref{Complex tridiagonal Hessenberg}).

The full Kostant-Toda lattice evolves on the set of full complex Hessenberg matrices
\begin{equation}
X = \left( \begin{array}{ccccc} * & 1 & 0 & \cdots & 0 \\ 
* & * & 1 & \cdots & 0 \\
\vdots& \vdots & \vdots & \ddots & \vdots \\
* & * & * & \cdots & 1  \\
* & * & * &\cdots  & *
\end{array} \right) \ \label{full Hessenberg matrix}
\end{equation}
with arbitrary complex entries below the diagonal.
The set of all such $X$ is denoted $\epsilon + \frak{b}_-$, where $\epsilon$ is the matrix with 1's on the superdiagonal and zeros elsewhere
and $\frak{b}_-$ is the set of lower triangular complex matrices. Note the decomposition
$\frak{sl}_n=\frak{b}_-\oplus \frak{n}_+$ where $\frak{n}_+$ is the set of strictly upper triangular matrices.

With respect to the symplectic structure on $\epsilon + \frak{b}_-$ (defined below), the Toda hierarchy 
(\ref{Hessenberg hierarchy}) with $X$ as in (\ref{full Hessenberg matrix}) is completely integrable on the generic leaves. 
The complete integrability is found in \cite{EFS}
by extending the techniques in \cite{DLNT:86} to $\epsilon + \frak{b}_-$.  
For complete integrability when $n > 3$, we need additional constants of motion independent of the eigenvalues of the initial matrix.  
These integrals, and the Casimirs (where the flows are trivial) are computed by a chopping construction on $\epsilon + \frak{b}_-$
that  creates matrices $\phi_k(X) \in GL(n-2k,{\mathbb C})$ for $0 \leq k \leq [\frac{(n-1)}{2}]$.  The coefficients of the polynomial
${\rm det}(\lambda - \phi_k(X))$ are constants of motion referred to as the $k$-chop integrals \cite{EFS};
they are
equivalent to the traces of the powers of $\phi_k(X)$.
The Hamiltonian system generated by an integral $I(X)$ is
\begin{equation}
\frac{d}{dt}X = [X,  \ (\nabla I(X))_{<0}]  \label{Toda equation for I} \ .
\end{equation}

The level sets of the $k$-chop integrals live on the leaves of a symplectic structure 
 on $\epsilon + \frak{b}_-$.  The symplectic structure can be defined as follows. Write
$\frak{sl}_n = \frak{n}_- \oplus \frak{b}_+ $
where $\frak{n}_-$ and $\frak{b}_+$ are the strictly lower triangular and the upper triangular subalgebras.
With a nondegenerate inner product $\langle A,B\rangle={\rm tr}(AB)$ on $\frak{sl}_n({\mathbb C})$,
we have an isomorphism $\frak{sl}_n\cong \frak{sl}_n^*$, where
$\frak{sl}_n^*=\frak{n}_-^*\oplus \frak{b}_+^*=\frak{b}_+^{\perp}\oplus \frak{n}_-^{\perp}$.
 With the isomorphisms
$\frak{b}_+^*\cong \frak{n}_-^{\perp}=\frak{b}_-$ and
$\frak{n}_-^*\cong \frak{b}_+^{\perp}=\frak{n}_+$,
we have
\[
\epsilon+\frak{b}_-\cong \frak{b}_+^*\,,
\]
which is the phase space of the full Kostant-Toda lattice. On $\frak{b}_+^*$,
the Lie-Poisson structure is the Kostant-Kirillov form, 
\[
\{f,g\}(X)=\langle X,[\Pi_{\frak{b}+}\nabla f,\Pi_{\frak{b}_+}\nabla g]\rangle\qquad\text{for}\quad X\in\frak{b}_+^*,
\]
which stratifies it into symplectic leaves \cite{EFS}.  


Consider the isospectral set $(\epsilon + \frak{b}_-)_{\Lambda}$ with fixed eigenvalues $\Lambda$.
Using (\ref{Kostantfactorization}), 
there is a unique lower unipotent
matrix $n\in N_-$ such that for $X \in (\epsilon + \frak{b}_-)_{\Lambda}$, $X = n C_{\Lambda} n^{-1}$, 
where $C_\Lambda$ is the companion matrix of $X$:
\begin{equation}\label{companion matrix}
C_{\Lambda}=\begin{pmatrix}
0        &  1       &   0      &  \cdots  &  0   \\
0        &   0       &  1      &\cdots    & 0  \\
\vdots&\vdots&\ddots&\ddots    &\vdots \\
0        &  0        &    \cdots &  0     &   1   \\
p_{n} & p_{n-1}&\cdots & p_2& 0
\end{pmatrix} \, ,
\end{equation}
where the $p_j$'s are the symmetric polynomials of the eigenvalues $\lambda_j$, that is,
\[
{\rm det}(\lambda I-X)=\prod_{j=1}^n(\lambda-\lambda_j)=\lambda^n-\sum_{j=2}^np_j\lambda^{n-j}\,.
\]  

The mapping
\begin{equation}\label{companion embedding}\begin{array}{cccccc}
c_{\Lambda} &:&  (\epsilon + \frak{b}_-)_{\Lambda} &\longrightarrow& SL(n, {\mathbb C})/B\\[1.5ex]
 & &  X &\longmapsto & n^{-1} \ \bmod B \end{array}
\end{equation}
is an embedding \cite{Kostant Whittaker}, referred to as the \emph{companion embedding}.
Its image is open and dense in the flag variety. 
Under this embedding,  
the $n-1$ flows of the 0-chop integrals  $\frac{1}{k} {\rm tr}X^k$
generate the action of the
centralizer of $C_{\Lambda}$ in $SL(n,{\mathbb C})$ (the group acts by multiplication on the left).  

When the eigenvalues  $\lambda_i$ are
distinct, $C_{\Lambda} = V \Lambda 
V^{-1}$, where  $\Lambda=\text{diag}(\lambda_1,\ldots,\lambda_n)$ and $V$ is a Vandermonde matrix $V=(\lambda_j^{i-1})$, and we have
$$X = n V \Lambda V^{-1} n^{-1}. $$
This gives an embedding 
\begin{equation}\label{Torus Embedding}\begin{array}{ccccc}
\Psi_{\Lambda} &: &(\epsilon + \frak{b}_-)_{\Lambda}& \longrightarrow&
SL(n,{\mathbb C})/B\\[1.5ex]
& &  X  &\longmapsto &V^{-1} n^{-1} \bmod B \ , \end{array} 
\end{equation}
under which the group generated by Hamiltonian flows of $H_k = (1/(k+1)) \mbox{tr}(X^{k+1})$
for $k = 1, ..., n-1$ is the maximal diagonal torus.
$\Psi_{\Lambda}$ is called
the {\it torus embedding}. 

When the values of the integrals are sufficiently generic (in particular, when the eigenvalues of each $k$-chop are distinct),
the flows of the $k$-chop integrals can be organized in the flag variety by the torus embedding as follows \cite{EFS}.
(The companion embedding gives a similar structure, but the torus embedding is more convenient since the group action is diagonal.)

Let 
$SL(m,{\mathbb C})/P$ denote the quotient of $SL(m,{\mathbb C})$ by the parabolic subgroup $P$ of $SL(m,{\mathbb C})$ whose 
entries below the diagonal in the first 
column and to the left of the diagonal in the last row are zero.
\cite{EFS} builds a tower of fibrations 
where the $k$-chop flows generate a level set of the $(k+1)$-chop integrals
in the partial flag variety $SL(n-2k,{\mathbb C})/P$ and the
$(k+1)$-flows act as a torus action along the fiber, $SL(n-2(k+1),{\mathbb C})/B$.
In the end, the closure of a level set of all the $k$-chop integrals in $SL(n, {\mathbb C})/B$ is realized as a product of closures of 
generic torus orbits in the product of partial flag varieties.
\begin{equation}
SL(n,{\mathbb C})/P \times SL(n-2, {\mathbb C})/P \times \cdots \times SL(n-2M, {\mathbb C})/P \label{product}
\end{equation}
where $M$ is largest $k$ for which there are $k$-chop integrals.


In \cite{GS:99}, Gekhtman and Shapiro generalize the full Kostant-Toda flows and
the $k$-chop construction of the integrals  to arbitrary simple Lie algebras, showing that the Toda flows on a 
generic coadjoint orbit in a simple Lie algebra $g$
are completely integrable.  A key observation in making this extension is that the 1-chop matrix 
$\phi_1(X)$ can be obtained as the middle $(n-2) \times (n-2)$ block of
$Ad_{\Gamma(X)}(X)$, where $\Gamma(X)$ is a special element of the Borel subgroup of $G$.
This allows one to use the adjoint action of a Borel subgroup, followed by a projection onto a subalgebra, to define the appropriate analog 
of the 1-chop matrix.

Finally, we note that full Kostant-Toda lattice has a symmetry of order two induced by the nontrivial automorphism of the Dynkin diagram of the Lie 
algebra $\frak{sl}_n({\mathbb C})$.  In terms of the matrices in $\epsilon + \frak{b}_-$, the involution is reflection along the anti-diagonal.
It is shown by Shipman in \cite{Symmetry} that this involution preserves all the $k$-chop integrals and thus defines an involution on each level set of the
constants of motion.  In the flag variety, the symmetry interchanges the two fixed points of the torus action that correspond to antipodal 
vertices of the moment polytope under the moment map (\ref{moment map}).

\subsection{Nongeneric full Kostant-Toda flows} \label{SSec:Nongeneric}

When eigenvalues of the initial matrix in $\epsilon + \frak{b}_-$ coincide, the torus embedding 
(\ref{Torus Embedding}) is not defined since each matrix 
in $\epsilon + \frak{b}_-$ has one Jordan block for each eigenvalue.  In the most degenerate case, when all eigenvalues are zero, \cite{PJM} uses the companion embedding (\ref{companion embedding}) 
to study the geometry of the flows.

If eigenvalues of each $k$-chop matrix $\phi_k(X)$ are distinct but one or more
eigenvalues of $\phi_j(X)$ and $\phi_{j+1}(X)$ coincide for some $j$, then
the torus orbits generated by the $k$-chop integrals in the product (\ref{product}) degenerate into unions of nongeneric orbits \cite{Nongeneric}. This 
is reflected in splittings of the moment polytopes of the partial flag varieties in (\ref{product}).

\begin{figure}[ht!]
\noindent
\begin{minipage}[c]{.35\linewidth}
\includegraphics[width=\textwidth]{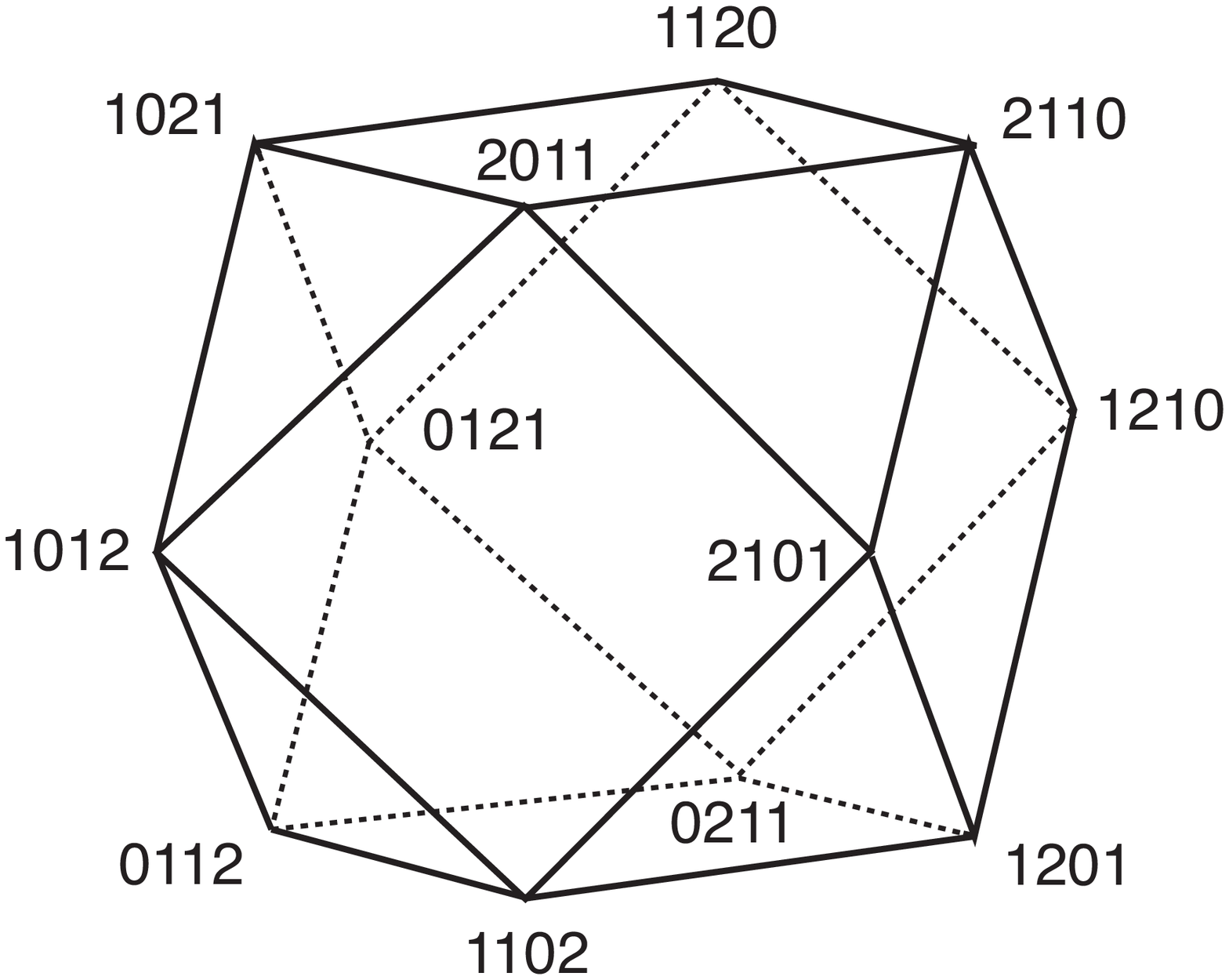}
\end{minipage} \hspace{.4cm}
\begin{minipage}[c]{.50\linewidth}
\includegraphics[width=\textwidth]{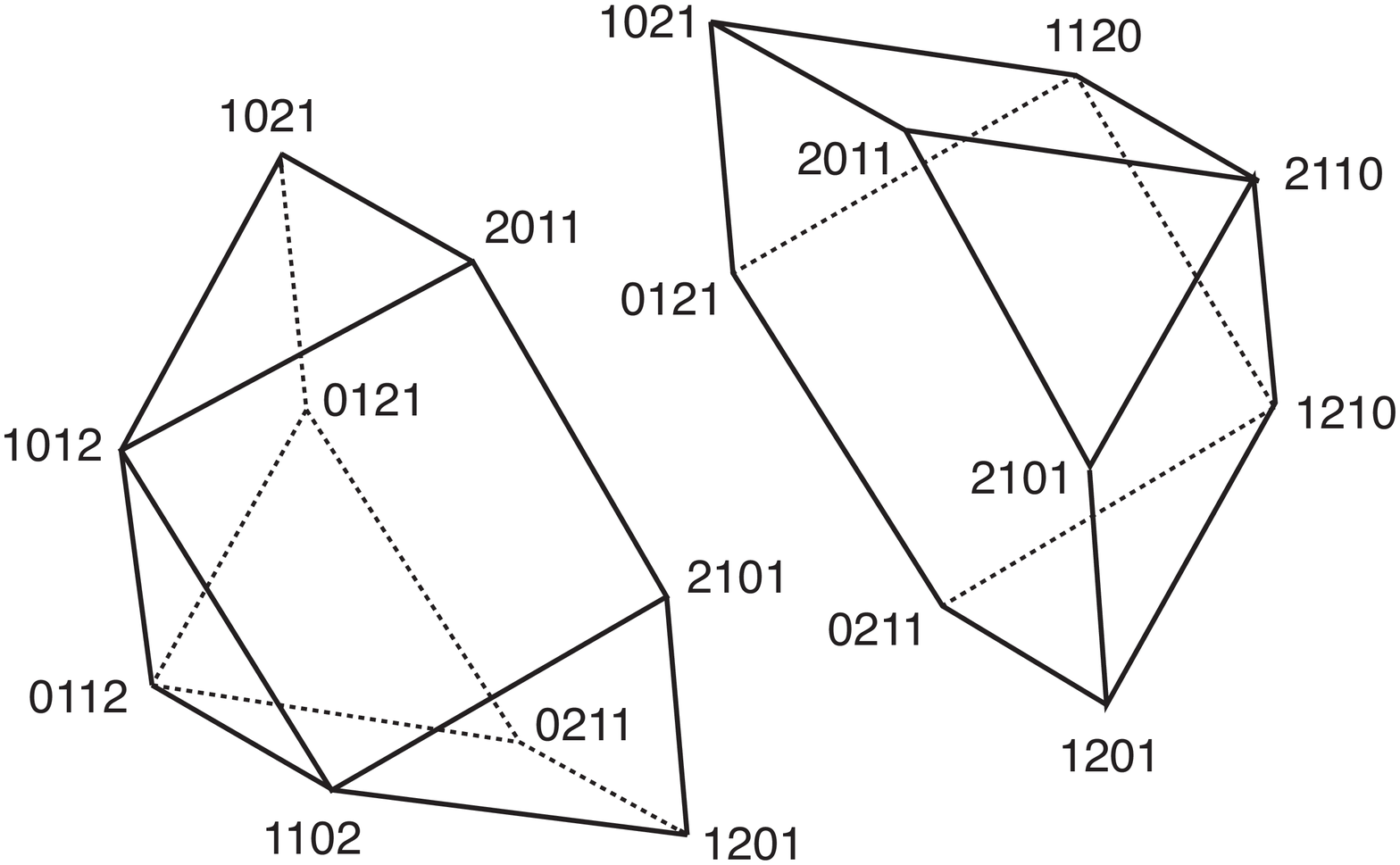}
\end{minipage}
\caption{Left: the moment polytope of $SL(4,{\mathbb C})/P$ where
the vertices are the weights ${\sf L}_i-{\sf L}_j$, 
e.g., $2110$ means ${\sf L}_1-{\sf L}_4=2{\sf L}_1+{\sf L}_2+{\sf L}_3$.
Right: complementary polytopes obtained
by splitting the moment polytope of $SL(4,{\mathbb C})/P$ along an interior hexagon.
}
\label{fig:sensAnalysis}
\end{figure}

From \cite{Nongeneric}, let 
${\mathcal F}$ be 
a variety in $SL(n,{\mathbb C})/P$ defined by fixing the values of the 1-chop integrals $I_{r1}$,
including the Casimir, where the values are chosen so that exactly one 
eigenvalue, say $\lambda_{i0}$, of $X$ is also an eigenvalue
of $\phi_1(X)$.
Then ${\mathcal F}$ is the union of the closures of two nongeneric
torus orbits such that the images of their closures under the moment map are obtained by splitting the moment polytope of $SL(n, \mathbb{C})/P$ 
along an interior face.
An example with $n=4$ is illustrated in Figure \ref{fig:sensAnalysis}.

This extends to degeneracies in $k$-chop flows for all $k$ \cite{Nongeneric}.
When a level set of the constants of motion is split into two or more nongeneric torus orbits, separatrices appear in the Toda flows that generate the torus action.   The faces along which the polytope is split are the images under
the moment map of lower-dimensional torus orbits (the separatrices) that form the interface between the nongeneric orbits of maximum dimension.

For $n=4$, Shipman in
\cite{Sl4} determines the monodromy of generic level sets around the singular fibers
in the fiber
bundle of level sets where the spectrum of the initial matrix is fixed 
and the single 1-chop integral $I$ is allowed to vary.  
The flow generated by $I$ produces a ${\mathbb C}^*$-bundle
with singular fibers over the values of $I$.  
The singularities occur at two types of coincidences: (1) at values of $I$ where an eigenvalue of the 1-chop matrix coincides with an eigenvalue of the original matrix and (2) at values of $I$ where the two eigenvalues of the 1-chop matrix coincide.   In a neighborhood of a singular
fiber of the first kind, the monodromy is characterized by a single twist of the noncompact cycle around the cylinder ${\mathbb C}^*$.  Near a singular fiber of the second kind, the monodromy creates two twists of the noncompact cycle.  This double twist appears in the simplest case when $n = 2$, around the level set where the two eigenvalues coincide \cite{Monodromy}.

\section{Other Extensions of the Toda Lattice}\label{Other Extensions}

In \cite{general matrix}, Kodama and Ye consider an iso-spectral deformation of an arbitrary diagonalizable matrix $L$.  The evolution equation is
\begin{equation} \label{general matrix}
\frac{d}{dt} L = [P, L] \qquad\text{with}\qquad
P  = (L)_{>0} - (L)_{<0} \,.
\end{equation}
The complete integrability of (\ref{general matrix}) is shown in \cite{general matrix} using inverse scattering; it generalizes
the method used in \cite{KM:96} to solve the full symmetric real Toda lattice.  The method yields an explicit solution to the initial-value
problem.  The general context of the flow (\ref{general matrix})
includes as special cases the Toda lattices on other classical Lie algebras in addition to
$\frak{sl}_n({\mathbb{R}})$, which is most closely associated with Toda's original system.  
In this regard, Bogoyavlensky in \cite{B:76} formulated the Toda lattice on the real split semisimple
Lie algebras, which are defined as follows (the formulation below is in the Hessenberg (or Kostant)
form, see also \cite{Guest,Perelomov}):
Let $\{h_{\alpha_i},e_{\pm\alpha_i}: i=1,\ldots,l\}$ be the Chevalley basis of the algebra $\frak g$ of rank $l$,
that is,
\[
[h_{\alpha_i},h_{\alpha_j}]=0,\quad [h_{\alpha_i},e_{\pm\alpha_j}]=\pm C_{ji}e_{\pm\alpha_j},\quad
[e_{\alpha_i},e_{-\alpha_j}]=\delta_{ij}h_{\alpha_j}\,,
\]
where $(C_{ij})_{1\le i,j\le l}$ is the Cartan matrix and $C_{ij}=\alpha_i(h_{\alpha_j})$.
Then the (nonperiodic) Toda lattice associated with the Lie algebra $g$ is governed by
the Lax equation
\begin{equation}\label{generalLax}
\frac{dL}{dt}=[A,L]\,,
\end{equation}
where $L$ is a Jacobi element of $\frak g$ and $A$ is the projection of $L$ onto $\frak{n}_-$, as
\begin{equation}\label{Bogo}\left\{\begin{array}{llll}
\displaystyle{L(t) = \sum_{i=1}^l \, f_i(t)\,h_{\alpha_i}+\sum_{i=1}^l \,(g_i(t)\,e_{-\alpha_i}+e_{\alpha_i})}\,,\\[2.0ex]
\displaystyle{A(t)=- \Pi_{\frak{n}_-}L(t)=-\sum_{i=1}^l\,g_i(t)\,e_{-\alpha_i}\,.}
\end{array}\right.
\end{equation}
The complete integrability is based on the existence of the Chevalley invariants of the algebra,
and the geometry of the isospectral variety has been discussed in terms of the representation
theory of Lie groups by Kostant in \cite{Kostant} for the cases
where $g_i$ are real positive, or complex. The general case for real $g_i$'s is studied
by Casian and Kodama \cite{CK:02, CK:02a}, which extends the results in
the $\frak{sl}_n(\Real)$ Toda lattice in the Hessenberg form (see Section \ref{Extended Hessenberg Toda}) to the Toda lattice for any real split semisimple Lie algebra.
 
The Lax equation (\ref{generalLax}) then gives 
\begin{align*}
\frac{df_i}{dt}=g_i, \qquad\text{and}\qquad
\frac{dg_i}{dt}=-\left(\sum_{j=1}^lC_{ij}f_j\right)\,g_i
\end{align*}
 from which the $\tau$-functions are defined as
\begin{equation}\label{general tau}
f_i(t)=\frac{d}{dt}\ln\tau_i(t),\qquad g_i(t)=g_i(0)\prod_{j=1}^l (\tau_j(t))^{-C_{ij}}\,.
\end{equation}
In the case of $\frak{g}=\frak{sl}_n(\Real)$, those equations are (\ref{g indefinite}).
Note here that the superdiagonal of $L(t)$ is ${\rm diag}(f_1-f_2,f_2-f_3,\ldots, f_{l}-f_{l+1})$
with $n=l+1$. Those extensions have been discussed by many authors (see for example
\cite{Guest, Perelomov}).
One should note that Bogoyavlensky in \cite{B:76} also formulates those Toda lattices
for affine Kac-Moody Lie algebras, and they give the periodic Toda lattice. There has been much 
progress in understanding these periodic Toda lattices, but we will not cover the subject in this paper
(see, for example \cite{AvM:80a, AvM:80b, Dub, RSTS, RS:94}).

From the viewpoint of Lie theory, the underlying structure of the integrable systems is
based on the Lie algebra splitting, e.g. $\frak{sl}_n=\frak{b}_-\oplus \frak{so}_n$ (the QR-factorization) for the symmetric
Toda lattice, and $\frak{sl}_n=\frak{b}_+\oplus \frak{n}_-$ (the LU decomposition) for the Hessenberg form of
Toda lattice. Then one can also consider the following form of the evolution equation,
\begin{equation}
\frac{d}{dt}L=[Q,\ L]  \qquad {\rm with}\quad Q=\Pi_{\mathfrak{g}_1}(L)\,, \label{Q general Lax}
\end{equation}
where $\mathfrak{g}_1$ is a subalgebra in the Lie algebra splitting $\frak{sl}_n=\mathfrak{g}_1\oplus\mathfrak{g}_2$.
In this regard, we mention here the following two interesting systems directly connecting to the Toda lattice:

\subsection{The Kac-van Moerbeke system} \cite{KM:75}:  We take $\frak{g}_1=\frak{so}_{2n}$, and consider the equation for $L\in \frak{so}_{2n}$.
Since $L^{2k-1}\in \frak{so}_{2n}$, the even flows are all trivial. Let $L$ be given by a tridiagonal form,
\[
L=\begin{pmatrix}
0   &   \alpha_1  &    0     &   \cdots    &    0    \\
-\alpha_1 &  0    &  \alpha_2 & \cdots  &   0  \\
\vdots  &  \vdots  &  \ddots   &   \cdots   & \cdots  \\
0          &    0        &    \cdots  &      0   &  \alpha_{2n-1} \\
0     &     0     &  \cdots   &  -\alpha_{2n-1}  &   0   \\
\end{pmatrix} \, \in \, \frak{so}_{2n}(\Real)
\]
Then the even flows are the Kac-van Moerbeke hierarchy, $\displaystyle{\frac{\partial L}{\partial t_{2j}}=[\Pi_{\frak{so}}(L^{2j}),L]}$,
where the first member of $t_2$-flow gives
\[
\frac{\partial \alpha_{k}}{\partial t_{2}}=\alpha_{k}(\alpha_{k-1}^2-\alpha_{k+1}^2)\,,\qquad k=1,\ldots,2n-1\,,
\]
with $\alpha_0=\alpha_{2n}=0$.
This system is equivalent to the symmetric Toda lattice which can be written as (\ref{Q general Lax})
for the square $L^2$. Note here that $L^2$ is a symmetric matrix given by
\[
L^2=T^{(1)}\otimes \begin{pmatrix}1 & 0\\0&0\end{pmatrix}\,+\, T^{(2)}\otimes \begin{pmatrix}0&0\\0&1\end{pmatrix}\,,
\]
where $T^{(i)}$, for $i=1,2$, are $n\times n$ symmetric tridiagonal matrices given by
\[
T^{(i)}=\begin{pmatrix}
b^{(i)}_1    & a^{(i)}_1 & 0 & \cdots & 0  \\
a^{(i)}_1 &  b_2^{(i)} & a_2^{(i)} & \cdots  & 0\\
\vdots    &   \vdots   &  \ddots   &  \ddots   &   \vdots  \\
0     &           0        &             \cdots  &     b_{n-1}^{(i)}    &   a_{n-1}^{(i)}\\
0     &           0       &          \cdots  &     a_{n-1}^{(i)} &b_{n}^{(i)}
\end{pmatrix}\,,
\] 
with $a^{(1)}_k = \alpha_{2k-1}
\alpha_{2k} $, $b^{(1)}_k = - \alpha_{2k-2}^2 - \alpha_{2k-1}^2 $, 
$a^{(2)}_k=\alpha_{2k}\alpha_{2k+1}$, and
$b^{(2)}_k=-\alpha_{2k-1}^2-\alpha_{2k}^2$ (see \cite{GHSZ:93}). Then
one can show that each $T^{(i)}$ gives the symmetric Toda lattice, that is,
the Kac-van Moerbeke hierarchy for $L^2$ matrix
splits into two Toda lattices,
\[ 
\frac{\partial T^{(i)}}{\partial t_{2j}} = [\Pi_{{so}}(T^{(i)})^{j},
  T^{(i)}]\,\qquad\text{for}\quad  i=1,2\,.
\]
The equations for $T^{(i)}$ are connected by the Miura-type transformation, with the functions
 $(a^{(i)}_k,b^{(i)}_k)$, through the Kac-van Moerbeke variables $\alpha_k$ (see \cite{GHSZ:93}).
 
\subsection{The Pfaff lattice for a symplectic matrix} \cite{AM:02, KP:07,KP:08}: The Pfaff lattice is defined in the same form
with $\frak{g}_1=\frak{sp}_{2n}$ and $L$ in the Hessenberg form with $2\times 2$ block structure.
In particular, we consider the case $L\in \frak{sp}_{2n}$ having the form,
\[
L=
\begin{pmatrix}
\begin{matrix} 0 & \sigma_1 \\ b_1 & 0 \end{matrix} &\vline&
\begin{matrix} 0  & 0 \\ a_1 & 0 \end{matrix} &\vline& \cdots &\vline& 0_2
\\ \hline\bigxstrut
\begin{matrix} 0 & 0 \\ a_1 & 0 \end{matrix} &\vline&
\begin{matrix} 0 & \sigma_2 \\ b_2 & 0 \end{matrix} &\vline&\cdots &\vline& 0_2
\\ \hline\bigxstrut
\vdots &\vline& \vdots &\vline& \ddots &\vline& \vdots
\\ \hline\bigxstrut
0_2 & \vline& 0_2 &\vline& \cdots &\vline &\begin{matrix} 0 & \sigma_n \\ b_n & 0\end{matrix}
\end{pmatrix} \, \in \,\frak{sp}_{2n}(\Real) \,,
\]
where $0_2$ is the $2\times 2$ zero matrix. The variables $(a_k,b_k)$ and $\sigma_k=\pm1$
are those in the indefinite Toda lattice in \eqref{Hessenberg equation} through  $f_k=\sigma_kb_k$ and $g_k=\sigma_k\sigma_{K=1}a_k^2$. It should be noted again that the odd members
are trivial (since $L^{2k-1}\in \frak{sp}_{2n}$), and the even members give the indefinite Toda
lattice hierarchy \cite{KP:08}. Here one should note that $L^2$ can be written as
\[
L^2=\tilde{L}^T\otimes\begin{pmatrix}1 & 0\\0&0\end{pmatrix} \, +\, 
\tilde{L}\otimes\begin{pmatrix}0&0\\0&1\end{pmatrix}\,,
\]
where $\tilde L=\tilde D^{-1}X\tilde D$ with $\tilde D=\text{diag}(1,\sigma_2a_1,\ldots,\sigma_na_{n-1} )$.
 Then one can show that 
the generator $Q_{2j}$ of the Lax equation is given by
\[
Q_{2j}=\Pi_{\frak{sp}}(L^{2j})=-\tilde{B}^T_{j}\otimes\begin{pmatrix}
1&0\\0&0\end{pmatrix}\,+\,\tilde{B}_j\otimes\begin{pmatrix}0&0\\0&1\end{pmatrix}\,,
\]
where $\tilde{B}_j=\frac{1}{2}[(\tilde{L}^j)_{>0}-(\tilde{L}^j)_{<0}]$.
Then the hierarchy $\frac{d}{dt}L=[Q_{2j},L]$ gives the indefinite Toda lattice hierarchy (see Section \ref{Extended Hessenberg Toda} and \cite{KY:96, KY:98}).

\section{The full Kostant-Toda lattice in real variables}\label{RealKT}

Here we consider the full Kostant-Toda hierarchy \eqref{Hessenberg hierarchy} in real variables,
where we write 
\begin{equation}\label{fKT}
\frac{\partial X}{\partial t_k}=[(X^k)_{\ge 0},\ X]\qquad\text{for}\quad k=1,2,\ldots,n-1.
\end{equation}
We let ${\bf t}:=(t_1,\ldots,t_{n-1})$ denote the multi-time variables for the flows in the hierarchy.
  As is the case of the complex full Kostant-Toda lattice, the solution space can be described by the flag variety $G/B$. Here we consider the asymptotic behavior of the solutions for the regular flows of the full Kostant-Toda hierarchy.
Those regular solutions are associated to points in the \emph{totally nonnegative} (tnn) flag variety, denoted by $(G/B)_{\ge 0}$.  Then we  discuss the moment map images of the regular flows of the full Kostant-Toda hierarchy.  This section is a brief review of \cite{KW:15} which provides
a geometric structure of the iso-spectral variety for the full Kostant-Toda flows including nongeneric cases.

\subsection{Totally nonnegative parts of flag varieties}\label{tnn}
We begin with a brief review of the \emph{tnn parts} of the flag variety,  $(G/B)_{\ge 0}$ where $G=SL(n,\Real)$.

For each $1 \leq i \leq n-1$ we have a homomorphism
$\phi_i:{SL}(2,\R)\to{SL}(n,\R)$ such that
\[
\phi_i\begin{pmatrix} a& b\\c&d\end{pmatrix}=
\begin{pmatrix}
1 &             &       &       &           &     \\
   &\ddots  &        &      &            &        \\
   &             &   a   &   b  &          &       \\
   &             &   c    &   d  &         &       \\
   &            &          &       &  \ddots  &     \\
   &            &          &       &              & 1
   \end{pmatrix} ~\in~ {SL}(n,\Real),
\]
that is, $\phi_i$ replaces a $2\times 2$ block of the identity matrix with $\begin{pmatrix} a&b\\c&d\end{pmatrix}$, where $a$ is at the $(i,i)$-entry.
We have $1$-parameter subgroups of $G$
defined by
\begin{equation}\label{pinning}
x_i(m) = \phi_i \left(
                   \begin{array}{cc}
                     1 & m \\ 0 & 1\\
                   \end{array} \right)  \text{ and }\ 
y_i(m) = \phi_i \left(
                   \begin{array}{cc}
                     1 & 0 \\ m & 1\\
                   \end{array} \right) ,\
\text{ where }m \in \Real.
\end{equation}
The simple reflections $s_i \in W=\Sym_n$ are given by
$s_i:= \dot{s_i} T$ where $\dot{s_i} :=
                 \phi_i \left(
                   \begin{array}{cc}
                     0 & -1 \\ 1 & 0\\
                   \end{array} \right)$,
and any $w \in W$ can be expressed as a product $w = s_{i_1} s_{i_2}
\dots s_{i_\ell}$ with $\ell=\ell(w)$ factors. Here $\ell(w)$ denotes the length of $w$.  We set $\dot{w} =
\dot{s}_{i_1} \dot{s}_{i_2} \dots \dot{s}_{i_\ell}$.

There are two opposite Bruhat decompositions of $G/B$:
\begin{equation*}
G/B=\bigsqcup_{w\in W} B \dot w B/B=\bigsqcup_{v\in W}
N \dot v B/B.
\end{equation*}
We define the intersection of opposite Bruhat cells
\begin{equation*}
\mathcal R_{v,w}:=(B\dot w B/B)\cap (N\dot v B/B),
\end{equation*}
which is nonempty
precisely when $v\le w$.
The strata $\mathcal R_{v,w}$ are often called \emph{Richardson varieties}.

Now we define the totally nonnegative part of the flag.
\begin{definition} \cite{Lusztig3}
The \emph{tnn
part $N_{\geq 0}$ of $N$} is defined to be the semigroup in
$N$ generated by the $y_i(p)$ for $p \in \Real_{\geq 0}$ in \eqref{pinning}.
The
\emph{tnn part 
$(G/B)_{\geq 0}$
of $G/B$} 
is defined by \begin{equation*}
(G/B)_{\geq 0} := \overline{ \{\,n  B ~|~ n \in N_{\geq 0}\, \} },
\end{equation*}
where the closure is taken inside $G/B$ in its real topology.
We sometimes refer to $(G/B)_{\geq 0}$ as the 
\emph{tnn flag variety}.
\end{definition}
Lusztig \cite{Lusztig3, Lusztig2} introduced a 
natural decomposition of $(G/B)_{\geq 0}$:
For $v, w \in W$ with $v \leq w$, let
\begin{equation*}
\mathcal R_{v,w}^{>0} := \mathcal R_{v,w} \cap (G/B)_{\geq 0}.
\end{equation*}
Then the tnn part of the flag variety $G/B$ has the decomposition,
\begin{equation}\label{decomposition}
(G/B)_{\ge 0}=\bigsqcup_{w\in W}\left(\bigsqcup_{v\le w}\mathcal{R}^{>0}_{v,w}\right).
\end{equation}

Let $\w:= s_{i_1}\dots s_{i_m}$ be a reduced expression for $w\in W$.
A  {\it subexpression} $\v$ of $\w$
is a word obtained from the reduced expression $\w$ by replacing some of
the factors with $1$. For example, consider a reduced expression in 
the symmetric group $\Sym_4$, say $s_3
s_2 s_1 s_3 s_2 s_3$.  Then $1\, s_2\, 1\, 1\, s_2\, s_3$ is a
subexpression of $s_3 s_2 s_1 s_3 s_2 s_3$.
Given a subexpression $\v$,
we set $v_{(k)}$ to be the product of the leftmost $k$
factors of $\v$, if $k \geq 1$, and $v_{(0)}=1$. To parametrize each component of $\mathcal{R}_{v,w}^{>0}$, we need the following definition of the subexpressions of $w$:

\begin{definition}\label{d:Js}\cite{Deodhar, MR}
Given a subexpression $\v$ of  $\w=
s_{i_1} s_{i_2} \dots s_{i_m}$, we define
\begin{align*}
J^{\circ}_\v &:=\{k\in\{1,\dotsc,m\}\ |\  v_{(k-1)}<v_{(k)}\},\\
J^{+}_\v\, &:=\{k\in\{1,\dotsc,m\}\ |\  v_{(k-1)}=v_{(k)}\},\\
J^{\bullet}_\v &:=\{k\in\{1,\dotsc,m\}\ |\  v_{(k-1)}>v_{(k)}\}.
\end{align*}

The subexpression  $\v$
is called {\it
nondecreasing} if $v_{(j-1)}\le v_{(j)}$ for all $j=1,\dotsc, m$,
e.g.\ if $J^{\bullet}_\v=\emptyset$.
It is called {\it distinguished}
if we have
$v_{(j)}\le v_{(j-1)}\hspace{2pt} s_{i_j}$ for all
$j\in\{1,\dotsc,m\}.$
In other words, if right multiplication by $s_{i_j}$ decreases the
length of $v_{(j-1)}$, then in a distinguished subexpression we
must have
$v_{(j)}=v_{(j-1)}s_{i_j}$.
Finally, 
 $\v$ is called a {\it positive distinguished subexpression}
(or a PDS for short) if
$v_{(j-1)}< v_{(j-1)}s_{i_j}$ for all
$j\in\{1,\dotsc,m\}$.
In other words, it is distinguished and nondecreasing.
\end{definition}

It is then quite important to note that given $v\le w$
and a reduced expression $\w$ for $w$,
there is a unique PDS $\v_+$ for $v$ contained in $\w$ \cite{MR, KW3}.
The following theorem then provides a parameterization of the tnn part of the flag variety.

\begin{theorem}\label{t:parameterization}\cite[Proposition 5.2, Theorem 11.3]{MR}
Choose a reduced expression $\w=s_{i_1} \dots s_{i_m}$ for $w$ with $\ell(w)=m$.
To $v \leq w$ we associate the unique 
PDS 
$\v_+$ for $v$ in $\w$.  Then $J^{\bullet}_{\v^+} = \emptyset$.
We define 
\begin{equation}\label{eq:G}
G_{\v_+,\w}^{>0}:=\left\{g= g_1 g_2\cdots g_m \left
|\begin{array}{ll}
 g_\ell= y_{i_\ell}(p_\ell)& \text{ if $\ell\in J^{+}_\v$,}\\
 g_\ell=\dot s_{i_\ell}& \text{ if $\ell\in J^{\circ}_\v$,}
 \end{array}\right. \right\},
\end{equation}
where each $p_\ell$ ranges over $\Real_{>0}$.
The set $G_{\v_+,\w}^{>0}$ lies in $N\dot{v}\cap B\dot{w}B$,
$G_{\v_+,\w}^{>0} \cong \Real_{>0}^{\ell(w)-\ell(v)}$,
and the map $g\mapsto g B$ defines an isomorphism 
\begin{align*}\label{e:parameterization+}
G_{\v_+,\w}^{>0}&\quad\overset\sim\longrightarrow\quad\mathcal R_{v,w}^{>0}.
\end{align*}
\end{theorem}

\subsubsection{The Grassmannian and its tnn part.}
The \emph{real Grassmannian} $Gr(k,n)$ is the space of all
$k$-dimensional subspaces of $\Real^n$.  An element of
$Gr(k,n)$ can be viewed as a full-rank $k\times n$ matrix $A$ modulo left
multiplication by nonsingular $k\times k$ matrices.  In other words, two
$k\times n$ matrices are equivalent, i.e. they 
represent the same point in $Gr(k,n)$, if and only if they
can be obtained from each other by row operations.  

Let $\binom{[n]}{k}$ be the set of all $k$-element subsets of $[n]:=\{1,\dots,n\}$.
For $I\in \binom{[n]}{k}$, let $\Delta_I(A)$
be the {\it Pl\"ucker coordinate}, that is, the maximal minor of the $k\times n$ matrix $A$ located in the column set $I$.
The map $A\mapsto (\Delta_I(A))$, where $I$ ranges over $\binom{[n]}{k}$,
induces the {\it Pl\"ucker embedding\/} $\Gr(k,n)\hookrightarrow \mathbb{RP}^{\binom{n}{k}-1}$.

Just as for the flag variety, one may identify the
Grassmannian with a homogeneous space. 
Let $P_k$ be the parabolic subgroup which fixes the 
$k$-dimensional subspace spanned by $e_1,\dots, e_k$.
(This is a block upper-triangular matrix containing $B$.)
Then we may identify $Gr(k,n)$ with the 
space of cosets $G/P_k$.

There is a natural projection 
$\pi_k:G/B \to Gr(k,n)$.
One may equivalently express this projection as the map
$\pi_k:G/B \to G/P_k$, where 
$\pi_k(gB) = gP_k$.
Abusing notation, 
we simply write $\pi_k(g)=A_k$   with $A_k\in Gr(k,n)\cong G/P_k$
instead of  $\pi_k(gB)=gP_k$.

Concretely, for $g\in G$, $\pi_k(g)$ is represented 
by the $k\times n$ matrix
$A_k$ consisting of the leftmost $k$ columns
of $g$, i.e. 
\begin{equation}\label{eq:pi}
g=\begin{pmatrix}
g_{1,1}&\cdots&g_{1,k}&\cdots &g_{1,n}\\
\vdots&\ddots&\vdots&\vdots&\vdots\\
g_{k,1}&\cdots&g_{k,k}&\cdots& g_{k,n}\\
\vdots&\vdots&\vdots&\vdots&\vdots\\
g_{n,1}&\cdots&g_{n,k}&\cdots& g_{n,n}
\end{pmatrix}\quad\longmapsto
\quad A_k=\begin{pmatrix}
g_{1,1}&\cdots&g_{k,1}&\cdots&g_{n,1}\\
\vdots&\ddots & \vdots&\vdots&\vdots\\
g_{1,k}&\cdots&g_{k,k}&\cdots&g_{n,k}
\end{pmatrix}.
\end{equation}
This is equivalent to the following formula using the Pl\"ucker embedding
into the projectivization of the wedge product space $\mathbb{P}(\bigwedge^k\R^n)
\cong\mathbb{RP}^{\binom{n}{k}-1}$ with the standard basis
$\{e_i:i=1,\ldots,n\}$,
\begin{equation}\label{standard}
g\cdot e_1\wedge\cdots\wedge e_k=\sum_{1\le i_1<\cdots<i_k\le n}\Delta_{i_1,\ldots,i_k}(A_k)\,e_{i_1}\wedge\cdots \wedge e_{i_k}.
\end{equation}
The Pl\"ucker coordinates $\Delta_{i_1,\ldots,i_k}(A_k)$ are then given by
\[
\Delta_{i_1,\ldots,i_k}(A_k)=\langle e_{i_1}\wedge\cdots\wedge e_{i_k},~g\cdot e_1\wedge\cdots\wedge e_k\rangle,
\]
where $\langle\cdot,\cdot\rangle$
is the usual inner product on $\bigwedge^k\R^n$. 

Now the tnn part of the Grassmannian is then defined as follows:
\begin{definition}\label{def:TNNGrass}
The \emph{tnn
part of the Grassmannian} $Gr(k,n)_{\geq 0}$
is the image $\pi_k ((G/B)_{\geq 0})$.
Equivalently, 
$Gr(k,n)_{\geq 0}$ is the subset of $Gr(k,n)$ such that all
Pl\"ucker coordinates are nonnegative.
\end{definition}

Let $W_k = \langle s_1,\dots,\hat{s}_{k},\dots,s_{n-1} \rangle$
be a parabolic subgroup of $W = \Sym_n$ obtained by deleting the 
transposition $s_k$ from the generating set.
Let $W^k$ denote the set of minimal-length
coset representatives of $W/W_k$.
Recall that a \emph{descent} of a permutation $z$
is a position $j$ such that $z(j)>z(j+1)$.
Then $W^k$ is the subset of permutations
which have at most one descent, and if it exists,
that descent must be in position $k$.

Rietsch in \cite{Rietsch} shows that the tnn part of the Grassmannian $Gr(k,n)_{\ge 0}$
has a cellular decomposition (cf. \eqref{decomposition}),
\begin{equation}\label{GrassCell} 
Gr(k,n)_{\geq 0}  
 = \bigsqcup_{w\in W^k} \bigsqcup_{v \leq w} \mathcal P_{v,w}^{>0}
\end{equation}
where $\mathcal P_{v,w}^{>0} = \pi_k(\mathcal R_{v,w}^{>0})$.

\begin{definition}
Let $M$ be an $n \times n$ matrix with real entries.  
Any determinant of a $k \times k$ submatrix (for $1 \leq k \leq n$) is called
a \emph{flag minor} if its set of columns is precisely $\{1,2,\dots,k\}$, 
the leftmost $k$ columns of $M$.  Let  $\Delta_{I_k}^k(M)$ denote the flag minor where $I_k=\{i_1,\ldots,i_k\}$ is the set of rows.
And we say that $M$ is \emph{flag nonnegative} 
if all of its flag minors
are nonnegative.  
\end{definition}

Note that the flag minors of $g \in G$  are precisely the Pl\"ucker 
coordinates of the projections of $gB$
to the various Grassmannians $\pi_k(gB)$
for $1 \leq k\leq n$.  Then, one can show \cite{KW:15} that
any $g\in G_{\v_+,\w}^{>0}$ is a flag nonnegative.  That is, the Pl\"ucker coordinates in \eqref{standard} are all nonnegative
when $A_k$ is given by the matrix  $g\in G_{\v_+,\w}^{>0}$.

For any $z\in W$ we define the ordered set
$z \cdot [k] = \{z(1), \dots, z(k)\}$.  (By ordered set, we mean that
we sort the elements of $z \cdot [k]$ according to their value.)
Then we have the following.

\begin{lemma}\label{minors}\cite{KW:15}
Let $v \leq w$ be elements in $W=\Sym_n$, and choose
$z\in \Sym_n$ arbitrarily.  Choose a reduced subexpression
$\w$ for $w$; this determines the PDS
$\v_+$ for $v$ in $\w$.
Choose  any $g\in G_{\v_+,\w}^{>0}$.  Then we have
\[
\Delta_{z\cdot [k]}^k(g) > 0\qquad  \text{ for}\quad  1\leq k \leq n  
\]
if and only if 
$$v \leq z \leq w.$$
\end{lemma}
This lemma is a key to determine the polytope structure of nongeneric flows of
the full Kostant-Toda hierarchy.

\subsection{Full Kostant-Toda flows with totally nonnegative initial data}\label{fKTSolutions}
Let us recall that the solution of the full Kostant-Toda can be found by the companion embedding
\eqref{companion embedding} with the factorization $\exp(\Theta_{X^\0}(\t))=n(\t)b(\t)$ where
$\Theta_{X^\0}(\t):=\sum_{j=1}^{n-1}(X^\0)^jt_j$ with the initial matrix $X^\0=X(\0)$.  
Then we have the solution $X(\t)=n^{-1}(\t)X^\0n(\t)$.  This can be stated in
the following diagram \cite{FH:91, CK:02, KW:15}:
\begin{equation}\label{companionE}
\begin{CD}
X^\0  @>c_{\Lambda}>> n_0\,B\\
@V Ad_{n(\t)^{-1}}VV @VVV \\
X(\t) @>c_{\Lambda}>>\quad\left\{
\begin{array}{lll}
     ~~ n_0\,n(\t)\,B \\[0.5ex]
      = n_0 \exp(\Theta_{X^\0}(\t))\,B\\[0.5ex]
     =\exp(\Theta_{C_{\Lambda}}(\t))\,n_0\,B
\end{array}\right.
\end{CD}
\end{equation}
where $X^\0=n_0^{-1}C_{\Lambda}n_0$.  
That is, the initial matrix $X^\0=X(\0)$ determines the element $n_0\in N$, and each full Kostant-Toda flow corresponds to an $\exp(\Theta_{C_{\Lambda}}(\t))$-orbit on 
the flag variety with the initial point $n_0B$.

We now associate to each matrix $g\in G^{>0}_{\v_+,\w}$
(representing a point of $\mathcal{R}_{v,w}^{>0}$) an initial matrix 
$X^\0$ for the full Kostant-Toda hierarchy.  
First we note that the $\tau$-functions for the hierarchy can be also found in the same form as in \eqref{indefinite tauk}, i.e.
\begin{equation}\label{fKT-tau}
\tau_k(\t)=\left[\exp\left(\Theta_{X^\0}(\t)\right)\right]_k\qquad\text{for}\qquad k=1,\ldots,n-1.
\end{equation}
We then express the $\tau$-functions with the 
initial matrix $X^\0$ in terms of $g$.

Recall that $C_\Lambda V=V\Lambda$ where $V$ is the Vandermonde matrix $V=(\lambda_j^{i-1})$, and 
that $[M]_k$ denotes the $k$th principal minor of the matrix $M$.
Then,
for each matrix $g\in G^{>0}_{\v_+,\w}$
we can associate an initial matrix 
$X^\0 \in \mathcal{F}_{\Lambda}$, 
defined by $X^\0 = n_0^{-1} C_{\Lambda} n_0$, where 
$n_0 \in N$ and $b_0 \in B$ are uniquely determined by 
the equation $Eg = n_0 b_0$ (this decomposition is true when $g\in G_{\v_+,\w}^{>0}$). 
Then, the $\tau$-functions for 
the full Kostant-Toda hierarchy
with initial matrix $X^\0$ are given by 
\begin{equation}\label{tauEg}
\tau_k(\mathbf{t})=[\exp(\Theta_{C_{\Lambda}}(\mathbf{t}))n_0]_k=d_k\left[V\exp\left(\Theta_{\Lambda}(\mathbf{t})\right)g\right]_k,
\end{equation}
where $d_k=[b_0^{-1}]_k$.

\begin{remark}
The formula
$\exp(\Theta_{C_{\Lambda}}(\mathbf{t}))n_0=E\exp(\Theta_{\Lambda}(\mathbf{t}))gb_0^{-1}$
 implies that the full Kostant-Toda flow gives a (noncompact) torus action on
the flag variety.  More precisely, the torus $(\R_{>0})^n$ acts by 
$\exp(\Theta_{\Lambda}(\mathbf{t}))$ on the basis vectors consisting of
the columns of the Vandermonde matrix $V$, that is, we have
$\exp(\Theta_{X^\0}(\mathbf{t}))n_0B=V\exp(\Theta_{\Lambda}(\mathbf{t}))gB$.
\end{remark}

Then using the Binet-Cauchy lemma to \eqref{tauEg} and $A_k=\pi_k(g)$,  the $\tau$-function can be written as
\begin{equation}\label{tau}
\tau_k(\mathbf{t})=d_k \sum_{I \in {[n] \choose k}} \Delta_I(A_k) E_I(\mathbf{t}),
\end{equation}
where $E_I(\t)$ for $I=\{i_1,\ldots,i_k\}$ is defined by
\begin{equation}\label{Ei}
E_I(\mathbf{t}):=
\prod_{\ell<m}(\lambda_{i_m}-\lambda_{i_\ell})\,\prod_{j\in I}^nE_j(\t)
\quad\text{with}\quad E_j(\t)=e^{\theta_{j}(\mathbf{t})}.
\end{equation}

Since $g\in G_{\v_+,\w}^{>0}$ is a flag nonnegative, i.e. $\Delta_{I_k}(A_k)\ge0$ for all $I_k\in\binom{[n]}{k}$, the $\tau$-function is sign-definite.  This implies that the full Kostant-Toda flow is complete for all $\t=(t_1,\ldots,t_{n-1})\in\R^{n-1}$, when the initial matrix $X^0$ comes from a point in $G_{\v_+,\w}^{>0}$.

\begin{remark}
The $\tau$-function in \eqref{tau} has the Wronskian structure, that is, if we define
the functions $\{f_1,\ldots,f_k\}$ by
\[
(f_1(\mathbf{t}),\ldots,f_k(\mathbf{t})):=(E_1(\mathbf{t}),\ldots,E_n(\mathbf{t}))\,A_k^T,
\]
then we have
\[
\tau_k(\mathbf{t})=d_k\,{\rm Wr}(f_1(\mathbf{t}),\ldots,f_k(\mathbf{t})),
\]
where the Wronskian is for the $t_1$-variable.  
Furthermore, if we identify the first three variables
as $t_1=x,~t_2=y$ and $t_3=t$ in \eqref{tau},
then we obtain the $\tau$-function for the KP equation \cite{KP:70}
which gives rise to soliton solutions of the KP equation from the Grassmannian $Gr(k,n)$ \cite{KW2} (see \cite{K:17} for a review of the KP solitons).
That is, $\tau_k$ is associated with a point of the Grassmannian $Gr(k,n)$. Then the set of $\tau$-functions $(\tau_1,\ldots,\tau_{n-1})$ is associated with a point of the flag variety, and the solution space of the full Kostant-Toda hierarchy is naturally given by the complete flag variety.
\end{remark}

\subsection{Asymptotic behavior of the full Kostant-Toda lattice}\label{fKTPolytopes}
Here we consider the asymptotics of the solution 
$X(\t)$ to the full Kostant-Toda lattice
 where $X(\0) = X^\0$ is the initial matrix associated with $g\in G_{\v_+,\w}^{>0}$, i.e.
 the tnn part of the flag variety.

Recall that we have a fixed order $\lambda_1 < \dots < \lambda_n$
on the eigenvalues, and that 
$z \cdot [k]$ denotes the ordered set $\{z(1), z(2),\dots, z(k)\}$.  
Since $A_k = \pi_k(g)$ and $g\in N\dot{v}\cap B\dot{w}B$ (by Theorem 
\ref{t:parameterization}),  the lexicographically maximal and minimal elements in $\mathcal{M}(A_k)$
are respectively given by $w \cdot [k]$ and $v \cdot [k]$.
Because of the order $\lambda_1 < \dots < \lambda_n$, we have
the following with $E_i(t)=e^{\theta_i(t)}$ in \eqref{Ei},
\begin{align*}
&E_1\ll E_2\ll \cdots \ll E_n,\quad\text{as}\quad t\to \infty, \\[0.5ex]
&E_1\gg E_2\gg \cdots\gg E_n,\quad\text{as}\quad t \to -\infty,
\end{align*}
This implies that 
each $\tau_k({t})$-function from \eqref{tau} 
has the following asymptotic behavior:
\[
\tau_k({t})~\longrightarrow~\left\{\begin{array}{lll}
E_{w \cdot [k]}({t})\quad&\text{as}\quad t\to\infty\\[1.0ex]
E_{v \cdot [k]}({t})\quad&\text{as}\quad t\to-\infty
\end{array}\right.
\]
Then the diagonal element $f_k(t)$ in $X(t)$  in
the form \eqref{g indefinite} can be calculated as
\[
f_k(t)=\frac{d}{dt}\ln\frac{\tau_k}{\tau_{k-1}}~\longrightarrow~\frac{d}{dt}\ln\frac{E_{w\cdot[k]}}{E_{w\cdot[k-1]}}=\frac{d}{dt}\ln E_{w(k)}=\lambda_{w(k)}\quad\text{as}~t\to\infty.
\]
This implies that $X(t)$ 
 approaches a fixed point of the full Kostant-Toda flow as $t\to\pm\infty$,
\[
X(t)~\longrightarrow~\left\{\begin{array}{lll}
\epsilon+\text{diag}(\lambda_{w(1)},\lambda_{w(2)},\dots,\lambda_{w(n)})\quad&\text{as}\quad t\to\infty\\[1.5ex]
\epsilon+\text{diag}(\lambda_{v(1)},\lambda_{v(2)},\dots,\lambda_{v(n)})\quad &\text{as}\quad t\to-\infty
\end{array}\right.
\]
This can be extended to the asymptotic properties for the full Kostant-Toda hierarchy as follows:
First note that for any permutation $z \in \Sym_n$, 
one can find a multi-time $\mathbf{c}=(c_1,\dots, 
c_{n-1})\in \R^{n-1}$
such that 
$E_{z(1)}(\mathbf{c}) > E_{z(2)}(\mathbf{c}) > \dots >
E_{z(n)}(\mathbf{c}).$  This can be shown by considering the functions
 $\ell_i: \R\times \R^{n-1} \to \R$, 
\[
\ell_i(t_0,\mathbf{t})= t_0+
\lambda_i t_1 + \lambda_i^2 t_2 + \dots + \lambda_i^{n-1} t_{n-1} = t_0+\theta_{i}(\mathbf{t}) =(t_0,t_1,\ldots,t_{n-1})\cdot V_i,
\]
where $V_i$ is the $i$-th column vector of the Vandermonde matrix $V$.  Then one can find a point 
$(t_0,\mathbf{c})$ such 
that $\ell_{z(1)}(t_0,\mathbf{c}) > \ell_{z(2)}(t_0,\mathbf{c}) > 
\dots > \ell_{z(n)}(t_0,\mathbf{c})$, which also implies
that $E_{z(1)}(\mathbf{c}) > E_{z(2)}(\mathbf{c}) > \dots > E_{z(n)}(\mathbf{c}).$

Now assume that $v \leq z \leq w$. Then recall Lemma \ref{minors} which says $\Delta_{z\cdot[k]}^k(g)>0$ for all $k=1,\ldots,n-1$.
Note that $E_{z\cdot[k]}(\mathbf{c})$ dominates the other exponentials
in the $\tau_k$-function \eqref{tau} at the point $\mathbf{c}$.  Then, in the direction $\mathbf{t}(s)=s\mathbf{c}$ with the limit $s\to\infty$, we have
\[
\tau_k(\mathbf{t}(s))~\approx~d_k\Delta_{z\cdot[k]}^k(g)E_{z\cdot[k]}(\mathbf{t}(s))\quad\text{as}\quad s\to\infty.
\]
Now using the formula of $f_k(\mathbf{t})$ in \eqref{g indefinite} with $t=t_1$, one can 
see that $f_k(\t(s)) \to \lambda_{z(k)}$ as $s\to\infty$, i.e. $X(\t(s))$ approaches the fixed point as $s\to\infty$, i.e.
\begin{equation}\label{eq:limit}
X(\mathbf{t}(s))~\longrightarrow~\epsilon+\text{diag}(\lambda_{z(1)},\ldots,\lambda_{z(n)})\quad
\text{as}\quad s\to\infty.
\end{equation}


\subsection{The moment polytope of the full Kostant-Toda lattice}

We now present the image of the moment map on the full Kostant-Toda flows coming from
the tnn flag variety and construct certain convex polytopes 
that generalize the permutohedron. 

Recall that $\mathsf{L}_i$ denotes a weight of 
the standard representation of $\mathfrak{sl}_n$, and $\frak{h}^*_\R$ represents
the dual of the Cartan subalgebra $\frak{h}_\R$, see \eqref{dualCartan}.
For $I = \{i_1,\dots,i_k\}$, we set 
$\mathsf{L}(I)
=\mathsf{L}_{i_1}+\mathsf{L}_{i_2}+\cdots+\mathsf{L}_{i_k} \in \mathfrak{h}_{\R}^*.$
The \emph{moment map} for the Grassmannian
$\mu_k:Gr(k,n)\to\mathfrak{h}_{\R}^*$ is defined by 
\begin{equation}\label{momentGrassmannian}
\mu_k(A_k):=\frac{\sum_{I\in\mathcal{M}(A_k)}\left|\Delta_{I}(A_k)\right|^2\mathsf{L}(I)}{\sum_{I\in\mathcal{M}(A_k)}\left|\Delta_I(A_k)\right|^2},
\end{equation}
see e.g. \cite{GS, GGMS, Nongeneric}.  

We recall the following fundamental result of
Gelfand-Goresky-MacPherson-Serganova \cite{GGMS} on the moment map
for the Grassmannian (which in turn uses the convexity theorem of 
Atiyah \cite{A:82} and Guillemin-Sternberg \cite{GuS}).

\begin{theorem} \cite[Section 2]{GGMS} \label{thm:convex}
If $A_k \in Gr(k,n)$ and we consider the action of the 
torus $(\C^*)^n$ on $Gr(k,n)$ (which rescales columns of the 
matrix representing $A_k$), then the closure of the image of the moment map
applied to the torus orbit of $A_k$ is a
convex polytope 
\begin{equation}\label{Gamma}
\Gamma_{\M(A_k)} = \conv\{\mathsf{L}(I) \ \mid \ 
\Delta_I(A_k) \neq 0 \text{ i.e. }I\in \M(A_k)\}
\end{equation}
called a 
\emph{matroid polytope}, whose vertices
correspond to the
fixed points of the action of the torus.  
\end{theorem}

\begin{remark}
In representation theory, this polytope is a weight polytope of the fundamental representation of $\mathfrak{sl}_n$ on $\bigwedge^kV$, where $V$ is
the standard representation.  
\end{remark}

It should be noted  that
if $A_k \in Gr(k,n)$ and we consider the action of the 
\emph{positive} torus $(\R_{>0})^n$ on $Gr(k,n)$, the conclusion of 
Theorem \ref{thm:convex} still holds.


The \emph{moment map} for the flag variety 
$\mu: G/B^+ \to \mathfrak{h}_{\R}^*$ in \eqref{moment map} can be written in the form,
\[
\mu(g):= \sum_{k=1}^{n-1}\mu_k(A_k),\qquad \text{ where }\quad A_k = \pi_k(g).
\]


We now compute the image of the moment map 
$\mu:G/B\to\mathfrak{h}^*_{\R}$ when applied to the full Kostant-Toda flow
$\exp(\Theta_{C_\Lambda}(\mathbf{t}))$
on the point $n_0B$ of the flag variety 
described in \eqref{companionE}.

First recall $C_{\Lambda}V=V\Lambda$ and $Vg=n_0b_0$.  Then we have
\begin{align*}
&\exp(\Theta_{C_\Lambda}(\mathbf{t}))n_0\cdot e_1\wedge\cdots\wedge e_k  =
V\,e^{\Theta_{\Lambda}(\mathbf{t})}\,gb_0^{-1} \cdot e_1\wedge\cdots\wedge e_k\\
=&\sum_{1\le i_1<\cdots<i_k\le n}Ve^{\Theta_{\Lambda}(\mathbf{t})}e_{i_1}\wedge
\cdots\wedge e_{i_k}\langle e_{i_1}\wedge\cdots\wedge e_{i_k}, g b_0^{-1}\cdot
e_1\wedge\cdots\wedge e_k\rangle\\
=&d_k \sum_{1\le i_1<\cdots<i_k\le n}Ve^{\Theta_{\Lambda}(\mathbf{t})}e_{i_1}\wedge
\cdots\wedge e_{i_k}\langle e_{i_1}\wedge\cdots\wedge e_{i_k}, g \cdot
e_1\wedge\cdots\wedge e_k\rangle\\
=&d_k\sum_{1\le i_1<\cdots<i_k\le n}\Delta_{i_1,\ldots,i_k}(A_k) Ve^{\Theta_{\Lambda}(\mathbf{t})}e_{i_1}\wedge\cdots\wedge e_{i_k}\\
=&d_k\sum_{1\le i_1<\cdots<i_k\le n}\Delta_{i_1,\ldots,i_k}(A_ke^{\Theta_{\Lambda}(\mathbf{t})})  V_{i_1}\wedge\cdots\wedge V_{i_k},
\end{align*} 
where  $d_k=[b_0^{-1}]_k$ and $V_{i}=Ve_i=(1,\lambda_i,\ldots,\lambda_i^{n-1})^T$.


We now define
$\varphi(\mathbf{t};g):=
\mu(\exp(\Theta_{C_\Lambda}(\mathbf{t}))n_0)$ and  
$\varphi_k(\mathbf{t};g):= 
\mu_k(\tilde\pi_k(\exp(\Theta_{C_\Lambda}(\mathbf{t}))n_0))=\mu_k(A_ke^{\Theta_{\Lambda}(\mathbf{t})})$ with $\pi_k(g)=A_k$. 
Then we have
\begin{align}\label{momentT}
\varphi(\mathbf{t};g)&=\sum_{k=1}^{n-1}\varphi_k(\mathbf{t};g)\qquad\text{with}\quad
\varphi_k(\mathbf{t};g)=\sum_{I\in\mathcal{M}(A_k)}\alpha_{I}^k(\mathbf{t};g)\,\mathsf{L}(I),\\ \nonumber
&\text{and} \qquad \alpha_I^k(\mathbf{t}; g)
=\frac{\left(\Delta_{I}(A_ke^{\Theta_{\Lambda}(\mathbf{t})})\right)^2}
{\sum_{J\in\mathcal{M}(A_k)}\left(\Delta_J(A_k e^{\Theta_{\Lambda}(\mathbf{t})})\right)^2}.
\end{align}
Note here that  $0<\alpha_{I}^k(\mathbf{t};g)<1$ and $\sum_{I\in\mathcal{M}(A_k)}\alpha_I^k(\mathbf{t};g)=1$ for each $k$.

\begin{definition}
We define the \emph{moment map image of the full Kostant-Toda flow} for  $g\in G_{\v,\w}^{>0}$ to be the  set 
\[
\QQ_g=\overline{\left\{\varphi(\mathbf{t};g)~|~\mathbf{t}\in\R^{n-1}\right\}}:=\overline{\bigcup_{\mathbf{t}\in\R^{n-1}}\varphi(\mathbf{t};g)}.
\]
Here the closure is taken 
using the usual topology of the Euclidian norm on $\mathfrak{h}_\R^*\cong\R^{n-1}$. 
\end{definition}

Then we can show \cite{KW:15} that 
for each $k$,
the image $\QQ^k_g:=\overline{\{\varphi_k(\mathbf{t};g)~|~\mathbf{t}\in\R^{n-1}\}}$
is the corresponding matroid polytope from \eqref{Gamma}, i.e. 
\[
\QQ^k_g=\Gamma_{\M(A_k)} \quad \text{ where }~A_k = \pi_k(g).
\]

Now  we have the following proposition \cite{KW:15}:
\begin{proposition}\label{cor:Minkowski}
Let $g\in G_{\v_+,\w}^{>0}$.
Then the  moment map image $\QQ_g$ of 
the full Kostant-Toda flow for $g$  is a
Minkowski sum of matroid polytopes.  
More specifically, for $A_k=\pi_k(g)$, $k=1,\ldots,n-1$, we have
$$\QQ_g = \sum_{k=1}^{n-1} 
\Gamma_{\M(A_k)}.
$$
\end{proposition}

We also define a certain polytope which sits inside the permutohedron.
\begin{definition}
Let $v$ and $w$ be two permutations in $\Sym_n$ such that $v \leq w$.
We define the \emph{Bruhat interval polytope} associated to $(v,w)$ to be the following convex hull:
\begin{equation*}
\mathsf{P}_{v,w}:=\CH\{\mathsf{L}_z \in\mathfrak{h}_{\R}^* \ \vert~ v \leq z \leq w\}.
\end{equation*}
In other words, this is the convex hull of all permutation vectors
corresponding to permutations $z$ lying in the Bruhat interval $[v,w]$.
In particular, 
if $w=w_0$ and $v=e$, then we have $\mathsf{P}_{e,w_0} = \Perm_n$.
(See \cite{TW:15} for the further discussion on the Bruhat interval polytopes.)
\end{definition}

Finally we have the following theorem for the moment polytope of the full Kostant-Toda flow \cite{KW:15}.
\begin{theorem}\label{thm:polytope}
Let $g\in G_{\v_+,\w}^{>0}$.
Then the  moment map image of 
the full Kostant-Toda flow for $g$   is the Bruhat interval polytope $\mathsf{P}_{v,w}$, i.e.
\[
\QQ_g = \mathsf{P}_{v,w}.
\]
\end{theorem}

Note that from Proposition \ref{cor:Minkowski} and Theorem 
\ref{thm:polytope}, we have the following remark:

\begin{remark}\label{cor:BIPM}
The Bruhat interval polytope $\mathsf{P}_{v,w}$ is a Minkowski
sum of matroid polytopes 
$$\mathsf{P}_{v,w} = \sum_{k=1}^{n-1} \Gamma_{\M_k}.$$
Here $\M_k$ is the matroid defining the cell 
of $Gr(k,n)_{\geq 0}$ that we obtain by projecting the 
cell $\mathcal{R}_{v,w}^{>0}$ of $(G/B)_{\geq 0}$ to 
$Gr(k,n)_{\geq 0}$. 
\end{remark}

We note that each weight vector $\mathsf{L}_{i_1,\ldots,i_n}$ defined in \eqref{weights} can be associated to the
ordered set of eigenvalues,
\[
\mathsf{L}_{i_1,\ldots,i_n}\quad\Longleftrightarrow\quad (\lambda_{i_1},\lambda_{i_2},\ldots,\lambda_{i_n}).
\]
This means that each vertex of the Bruhat interval polytope can be labeled by the ordered set of eigenvalues. For example, the highest weight for the permutohedron of \eqref{perm} is given by
\[
\mathsf{L}_{1,2,\ldots,n}=\sum_{k=1}^n(n-k)\mathsf{L}_k\quad\Longleftrightarrow\quad(\lambda_1,\lambda_2,\ldots,\lambda_n).
\]
which corresponds to the asymptotic form of $\text{diag}(L)$ with $v=e$ for $t\to-\infty$.  The permutohedron  $\mathsf{P}_{e,w_0}$ with the longest element $w_0$ for $SL(4,\R)/B$ is illustrated in Fig. \ref{fig:Mpoly} (Left).

\begin{example}
Consider the $\mathfrak{sl}_4(\R)$ full Kostant-Toda hierarchy. We take 
\[
w=s_2s_3s_2s_1 \text{ and } v=s_3,
\]
which gives
\[
w\cdot(1,2,3,4)=(4,1,3,2) \text{ and } v\cdot(1,2,3,4)=(1,2,4,3).
\]
There are eight permutations $z$ satisfying
$v\le z\le w$, i.e.
\[
v=s_3,\quad s_3s_2,\quad s_2s_3,\quad s_3s_1,\quad s_3s_2s_1,\quad s_2s_3s_1,\quad s_2s_3s_2,\quad w=s_2s_3s_2s_1.
\]
We illustrate the moment polytopes in Figure \ref{fig:Mpoly}.  The vertices are labeled by the index set ``$i_1i_2i_3i_4$'' of the eigenvalues $(\lambda_{i_1},\lambda_{i_2},\lambda_{i_3},\lambda_{i_4})$. The vertex with the white circle indicates the asymptotic form of $\text{diag}(L)$ for
$t\to-\infty$ (i.e. $(1,2,4,3)=v\cdot(1,2,3,4)$ in the right figure), and the black one indicates the asymptotic form for $t\to\infty$ (i.e. $(4,1,3,2)=w\cdot(1,2,3,4)$ in the right figure).
\begin{figure}[h]
\centering
\includegraphics[height=5cm]{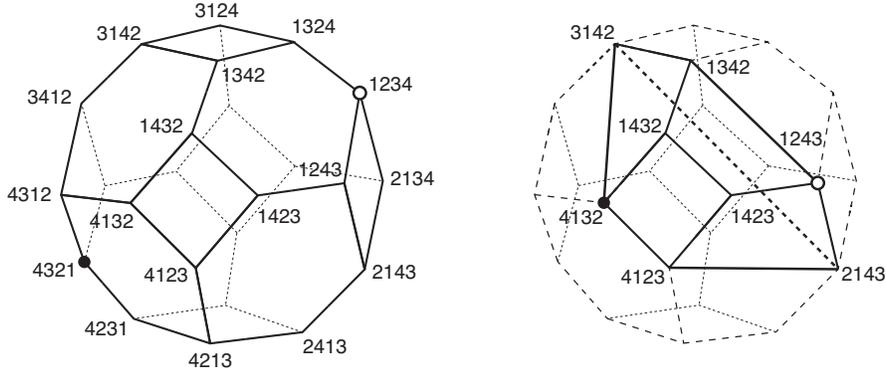}
\caption{Some moment polytopes from the
 $\mathfrak{sl}_4(\R)$ full Kostant-Toda hierachy.  Each 4-digit number represents the order of
 the eigenvalues, e.g. $2413=(\lambda_2,\lambda_4,\lambda_1,\lambda_3)$. Left: the
permutohedron $\mathsf{P}_{e,w_0}=\Perm_4$ with $e=1234$ and $w_0=4321$.  Right: the 
Bruhat interval polytope $\mathsf{P}_{v,w}$ with $w=s_2s_3s_2 s_1$ and $v=s_3$,
equivalently, they are represented as $w=4132$ and $v=1243$.}
\label{fig:Mpoly}
\end{figure}
\end{example}

\begin{remark}
The moment polytopes of the full symmetric Toda hierarchy can be shown to have the same structure as these of the full Kostant-Toda hierarchy on the tnn flag variety (see \cite{BG:98, CSS:14, KW:15}).
\end{remark}


\section{The Toda lattice and integral cohomology of real flag varieties} \label{Toda-cohomology connections}
Here we explain how one can find the integral cohomology of real flag variety
from the isospectral variety of the Toda lattice (this is an invitation to the papers \cite{CK:06,CK:07}).
We consider the Toda lattice hierarchy (\ref{Hessenberg hierarchy}) on the {\it real} split semi-simple Lie algebra
$\frak{sl}_n(\Real)$, and assume $X\in\frak{sl}_n(\Real)$ to be a generic element in the tridiagonal
Hessenberg form of \eqref{Hessenberg}, that is, it has all real and distinct eigenvalues (see \cite{CK:06,CK:07}, for the general case associated with real split semisimple Lie algebra).

\subsection{Integral cohomology of $G/B$}
We begin with a brief summary of the cohomology of $G/B$ as a background for
the next section where we explain how one gets the cohomology of $G/B$ from
the isospectral variety of the Toda lattice associated with real split semisimple Lie group $G$.

Let us first recall the Bruhat decomposition of $G/B$,
\[
G/B=\bigsqcup_{w\in W}\Omega^\circ_w \qquad {\rm with}\quad \Omega^\circ_w=NwB/B\,.
\]
Each Bruhat (or Schubert) cell $\Omega^\circ_w$ is labeled by the element $w\in W$ and ${\rm codim}\,(\Omega_w)=l(w)$ where $l(w)$ represents the length of $w$. 
Let $\sigma_w$ denote the Schubert class associated to the Schubert variety $\Omega_w=\cup_{w\le w'}\Omega^\circ_{w'}$.  Here the Bruhat order is defined as $w\le w'$ iff $\Omega_w\supset\Omega_{w'}$.
Then we can define the chain complex,
\[
\mathcal{C}^*=\bigoplus_{k=0}^{l(w_0)}\mathcal{C}^k\qquad{\rm with}\quad  \mathcal{C}^k=\sum_{l(w)=k}\mathbb{Z}\,\sigma_w\,,
\]
where $w_0$ is the longest element of $W$, and the coboundary operators $\delta_k:\mathcal{C}^k\to\mathcal{C}^{k+1}$  is given by
\[
\delta_k(\sigma_w)=\sum_{l(w')=k+1}[w:w']\,\sigma_{w'}\,,
\]
where $[w:w']$ is the incidence number associated with $\sigma_w\overset{\delta_k}{\longrightarrow}\sigma_{w'}$.
It has been known (see \cite{K:95, CS:99}) that the incidence number is either $0$ or $\pm2$ for the real flag variety
$G/B$ of real split semi-simple Lie group $G$. Then the cohomology
of $G/B$ can be calculated from the incidence graph $\mathcal{G}_{G/B}$ defined as follows:
\begin{definition}\label{incidence graph}
The incidence graph $\mathcal{G}_{G/B}$ consists of
the vertices labeled by $w\in W$ and the edges $\Rightarrow$ defined by
\[
w\,\Rightarrow\, w' \quad {\rm iff}\quad \left\{
\begin{array}{lllll}
{\rm (i)}~~ w\le w' \\[0.4ex]
{\rm (ii)}~~ l(w')=l(w)+1 \\[0.4ex]
{\rm (iii)}~~ [w:w']\ne 0
\end{array}
\right.
\]
The incidence number for each edge is either $0$ or $\pm 2$ (see \cite{CS:99}). The integral cohomology is then calculated from the graph.
\end{definition}
\begin{example}
In the case of $G=SL(3,\Real)$, the incidence graph and the integral cohomology are given by
\[
\begin{matrix} 
{}                     &{e}& {}                \\[1.5ex]
[1]                     &{} & [2]                   \\
 \Downarrow&{}&\Downarrow  \\
  [12]               &{}& [21]                  \\[1.5ex]
   {}&{[121]}&{ }                      \\
    \end{matrix} \hskip1.5cm\text{and}\hskip1.5cm
    \left\{ \begin{array}{lllll}
H^0(G/B,\mathbb Z )&= &\mathbb Z   \\[0.4ex]
H^1(G/B,\mathbb Z )&=& 0               \\[0.4ex]
H^2(G/B,\mathbb Z )&=& \mathbb{Z}_2\oplus\mathbb{Z}_2  \\[0.4ex]
H^3(G/B,\mathbb Z)&=&\mathbb Z \\
\end{array}\right. 
\] 
Here the Schubert classes are denoted by $[ij]$ for $w=s_is_j$, e.g. $\delta_1[2]=\pm2 [21]$ for
$\delta_1\sigma_{s_2}=\pm2\sigma_{s_2s_1}$.

\begin{figure}[t!]
\centering
\includegraphics[scale=0.46]{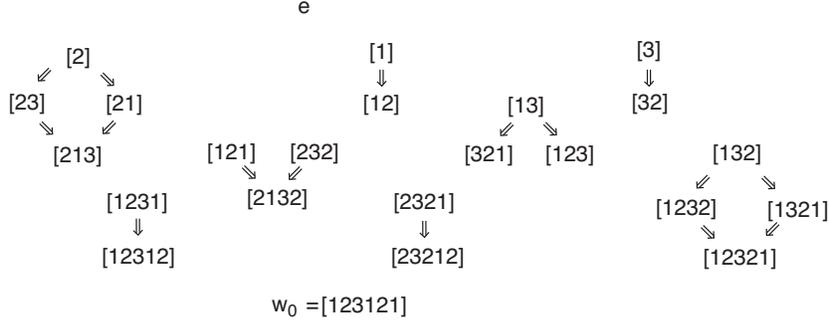}
\caption{The incidence graph $\mathcal{G}_{G/B}$ for the real flag variety $SL(4,\Real)/B$.
The Schubert class $\sigma_w$ is denoted by $[ij\ldots k]$ for $w=s_is_j\ldots s_k$.
$w_0=[123121]$ is the longest element of $W=\Sym_4$.
The incidence numbers associated with the edges $\Rightarrow$ are $\pm 2$ (see also Example 8.1
in \cite{CS:99}).}
\label{fig:A3incidence}
\end{figure}
In Figure \ref{fig:A3incidence}, we show the incidence graph for $G/B$ with $G=SL(4,\Real)$,
  from which one can compute the integral cohomology \cite{CS:99} as
\[
\left\{ \begin{array}{lllll}
H^0(G/B,\mathbb Z )&= &\mathbb Z   \\[0.4ex]
H^1(G/B,\mathbb Z )&=& 0               \\[0.4ex]
H^2(G/B,\mathbb Z )&=& \mathbb{Z}_2\oplus\mathbb{Z}_2\oplus\mathbb{Z}_2  \\[0.4ex]
H^3(G/B,\mathbb Z)&=&\mathbb{Z}\oplus\mathbb{Z}\oplus\mathbb{Z}_2\oplus\mathbb{Z}_2 \\[0.4ex]
H^4(G/B,\mathbb Z)&=&\mathbb{Z}_2\oplus\mathbb{Z}_2\\[0.4ex]
H^5(G/B,\mathbb{Z})&=&\mathbb{Z}_2\oplus\mathbb{Z}_2\oplus\mathbb{Z}_2\\[0.4ex]
H^6(G/B,\mathbb{Z})&=&\mathbb{Z}
\end{array}\right. 
\]

\end{example}

The incidence graph for the general case of real split semisimple $G$ can be found in \cite{CS:99}.
Then the integral cohomology of $G/B$ can be computed from the incidence graph with the incidence
numbers $[w:w']$ being $0$ or $\pm2$.

For the rational cohomology, we have 
\begin{equation}\label{KG relation}
H^*(G/B,\mathbb{Q})=H^*(K/T,\mathbb{Q})=H^*(K,\mathbb{Q})\,,
\end{equation}
where $K$ is the maximal compact subgroup of $G$, and $T$ is the maximal torus of $G$. e.g. for $G=SL(n,\Real)$, $K=SO(n)$, and 
$T={\rm diag}(\pm1,\ldots,\pm1)$ (see Proposition 6.3 in \cite{CK:07}).
It is well known that the cohomology ring $H^*(K,\mathbb{Q})$ of compact connected group $K$ of rank $l$ is given by the exterior product algebra,
\[
H^*(K,\mathbb{Q})=\bigwedge{}_{\mathbb{Q}}\{x_{m_1},x_{m_2},\ldots, x_{m_l}\}\,,
\]
where $\{x_{m_1},\ldots,x_{m_l}\}$ are the generators of the exterior product representation with
${\rm deg}(x_{m_i})=m_i$ (odd) for $i=1,\ldots,l$ and $m_1+\cdots+m_l={\rm dim}(K)$ (see for example \cite{C:89}).
In the case of $K=SO(n)$, we have,
\begin{itemize}
\item[(a)] for $n=2m+1$,
\[H^*(SO(2m+1),\mathbb{Q})=\bigwedge{}_{\mathbb{Q}}(x_3,x_7,\ldots,x_{4m-1})\]
\item[(b)] for $n=2m$,
\[H^*(SO(2m),\mathbb{Q})=\bigwedge{}_{\mathbb{Q}}(x_3,x_7,\ldots, x_{4m-5},y_{2m-1})\,.
\]
Note here that the generators include the additional $y_{2m-1}$. For example, $H^*(SO(4),\mathbb{Q})$ is generated by two elements $x_3,y_3$ of the same degree, $V=\mathbb{Q} x_3+\mathbb{Q} y_3$ and $\wedge^2V=\mathbb{Q} x_3\wedge y_3$.
\end{itemize}

We also note that the number of points on the finite Chevalley group $K(\mathbb{F}_q)$ of the
compact connected group $K$ is given by certain polynomial of $q$. Here $\mathbb{F}_q$ is a finite
field with $q$ elements. Although this polynomial can be computed by using the Lefschetz fixed point theorem for the Frobenius map $\Phi:K(\overline{\mathbb{F}}_q)\to K(\overline{\mathbb{F}}_q),\, 
x\mapsto x^q$, 
we here give an elementary calculation to find those polynomials for $K=SO(n)$
(see also \cite{CK:06}). As we show in the next section, those polynomials are also related to
the indefinite Toda lattice of Section \ref{Extended Hessenberg Toda}  \cite{KY:96, KY:98}.

Let us first assume that $q$ is a power of a prime number $p\ne 2$, such that in $\mathbb{F}_q$
the polynomial $x^2+1$ is not irreducible, i.e. $\sqrt{-1}\in\mathbb{F}_q$. Then we have
the following results for $|S^n(\mathbb{F}_q)|$, the number of $\mathbb{F}_q$ points on $S^n$:
\[
|S^n(\mathbb{F}_q)|=\left\{\begin{array}{lllll}
q^{m-1}(q^m-1)\quad & {\rm if}\quad & n=2m-1, \\[1.0ex]
q^m(q^m+1) & {\rm if}  &  n=2m.
\end{array}\right.
\]
This can be shown as follows:
Let us first consider the case $n=1$, i.e.
 \[
 S^1(\mathbb{F}_q)=\{(x,y)\in\mathbb{F}_q^2:x^2+y^2=1\}\,.
 \]
 Then using the formulae for the stereographic projection; $x=\frac{2u}{u^2+1},
 y=\frac{u^2-1}{u^2+1}$ with $y\ne 1$ and $\{u\in \mathbb{F}_q:u^2+1\ne0\}$. 
 Since $\sqrt{-1}\in\mathbb{F}_q$,
 we have $2$ points in $\{u^2+1=0\}$. Counting the point $(0,1)$, the north pole, we have
 $|S^{1}(\mathbb{F}_q)|=q-2+1=q-1$. Now consider the case $n=2$, we have
 $x=\frac{2u_1}{u_1^2+u_2^2+1}, y=\frac{2u_2}{u_1^2+u_2^2+1}, z=\frac{u_1^2+u_2^2-1}{u_1^2+u_2^2+1}$ with $z\ne 1$ and $\{(u_1,u_2)\in \mathbb{F}_q^2: u_1^2+u_2^2+1\ne 0\}$.
 This gives $q^2-(q-1)$ points (note $(q-1)$ is the number of points in $u_1^2+u_2^2+1=0$).
 We now add the points of the north pole $(x,y,1)$ with $x^2+y^2=0$. This gives $2(q-1)+1$,
 where $2(q-1)$ for $x=\pm\sqrt{-1}y\ne 0$ and $1$ for $(0,0,1)$. Then we have
 $|S^2(\mathbb{F}_q)|=q^2-(q-1)+2(q-1)+1=q(q+1)$. Using the induction, one can show
 that the number of points in the north pole is given by
 $|\{(x_1,\ldots, x_{2m-1})\in \mathbb{F}_q^{2m-1}:x_1^2+\cdots+x_{2m-1}^2=0\}|=q^{2m-2}$ and
 $|\{x_1,\ldots,x_{2m})\in \mathbb{F}_q^{2m}:x_1^2+\cdots+x_{2m}^2=0\}|=q^{2m-1}+q^m-q^{m-1}$.
 Then one can obtain the above formulae for $|S^n(\mathbb{F}_q)|$.

We can now find the number of $\mathbb{F}_q$ points
of finite Chevalley group $SO(n,\mathbb{F}_q)$:
First recall that $SO(n+1)/SO(n)\cong S^n$. Then we obtain
\[
|SO(n,\mathbb{F}_q)|=\prod_{k=1}^n |S^{n-k}(\mathbb{F}_q)|\,,
\]
which leads to the results \cite{C:89}:
\begin{itemize}
\item[(a)] For $n=2m$,
\[
|SO(2m,\mathbb{F}_q)|=2q^{m(m-1)}(q^2-1)(q^4-1)\cdots (q^{2m-2}-1)(q^m-1)\,.
\]
\item[(b)] For $n=2m+1$,
\[
|SO(2m+1,\mathbb{F}_q)|=2q^{m^2}(q^2-1)(q^4-1)\cdots (q^{2m}-1)\,.
\]
\end{itemize}

In general,  the number of $\mathbb{F}_q$ points on the compact group $K$ can be
expressed by (see e.g. \cite{C:89})
\begin{equation}\label{Fq points} 
\left|K(\mathbb{F}_q)\right|=q^r\,p(q)\qquad {\rm with}\quad p(q)=\prod_{i=1}^l\,(q^{d_i}-1)\,,
\end{equation}
where $d_i$'s are degree of basic Weyl group 
invariant polynomials for $K$ given by $d_i=(m_i+1)/2$,
and $r={\rm dim}(K)-{\rm deg}(p(q))$. 
In the next section, we show that those polynomials can be reproduced by
counting the blow-ups in the solution of the indefinite Toda lattice 
(see \cite{CK:06} for the general case).


\subsection{Blow-ups of the indefinite Toda lattice on $G$ and the cohomology of $G/B$}\label{blow-ups} 
Now we show how to obtain the cohomology of $G/B$ from the moment polytope of
the indefinite Toda lattice of Section \ref{Extended Hessenberg Toda}.

First note that the $\tau$-functions can change their signs if some (but not all) of $\sigma_i$'s are negative.
This can be seen from (\ref{indefinite fg}), and implies that the solution blows up for some time $t_1=\bar{t}_1$, (see also (\ref{g indefinite})). The explicit form of the $\tau$-functions
can be obtained from (\ref{indefinite tauk}), and they are expressed by (see also Proposition 3.1 in \cite{KY:98}),
\begin{equation}\label{sign tau}
\tau_k(t)=\sum_{1\le j_1<\cdots<j_k\le n}\sigma_{j_1}\cdots \sigma_{j_k}\,K(j_1,\ldots,j_k)\exp\left(\sum_{i=1}^k\lambda_{j_i}t\right)\,,
\end{equation}
where $K(j_1,\ldots,j_k)$ are positive and given by
\[
K(j_1,\ldots,j_k)=\left(\varphi^0(\lambda_{j_1})\cdots\varphi^0(\lambda_{j_k})\right)^2\left|
\begin{matrix}
1          & \cdots    & 1 \\
\vdots & \ddots    & \vdots \\
\lambda_{j_1}^{k-1} & \cdots & \lambda_{j_k}^{k-1}\\
\end{matrix}\right|^2\,>0\,.
\]
As the simplest case, let us consider the $\frak{sl}_2(\Real)$ indefinite Toda lattice: We have one $\tau$-function, 
\[
\tau_1(t)=\sigma_1\rho_1e^{\lambda_1t}+\sigma_2\rho_2e^{\lambda_2t}\,.
\]
If $\sigma_1\sigma_2=-1$, $\tau_1(t)$ has zero at a time $t=\frac{1}{\lambda_2-\lambda_1}\ln(\frac{\rho_1}{\rho_2})$, that is, we have a blow-up in the solution.
The image of the moment map $\mu(\tau_1)$ is given by a line segment whose end points correspond 
to the weights ${\sf L}_1$ and ${\sf L}_2=-{\sf L}_1$. Although the dynamics are so different in the cases $\sigma_1\sigma_2 > 0$ and $\sigma_1\sigma_2 < 0$,
the moment polytope (a line segment) is independent of the signs of the $\sigma_i$'s.  
Notice that $\sigma_1\sigma_2={\rm sgn}(g_1)$,
and in general, if ${\rm sgn}(g_k)<0$ for some $k$, then the solution blows up sometime in $\Real$.

In order to find the general pattern of the sign changes in $(g_1(t),\ldots,g_{n-1}(t))$ of the matrix $X$ in (\ref{Hessenberg}) with \eqref{indefinite fg},
we first recall
that the isospectral variety is characterized by the moment polytope $\mathcal{M}_{\epsilon}$ whose vertices are given by
the orbit of Weyl group action.  Here the set of signs $\epsilon=(\epsilon_1,\ldots,\epsilon_{n-1})$
is defined by the signs of $g_i$ for $t\to -\infty$.
From the ordering $\lambda_1<\cdots<\lambda_n$, 
we first see that  $\tau_k(t)\approx \sigma_1\cdots \sigma_k K(1,\ldots,k)\exp((\lambda_1+\cdots+\lambda_k)t)$.
Then from the definition of $g_k(t)$ in (\ref{g indefinite}), i.e. $g_k=\tau_{k-1}\tau_{k+1}/\tau_k^2$,
the sign of
$g_k(t)$ for $t\to-\infty$ is given by
\[
\epsilon_k={\rm sgn}(g_k)=\sigma_k\sigma_{k+1}\, \quad {\rm for}\quad k=1,\ldots, n-1\,.
\]  
Then from the moment map (\ref{moment map}), one notes that the moment polytope given as the image of the moment map $\mu(\mathcal{M}_{\epsilon})$
 is independent of the sign set $\epsilon$.
However the dynamics of the Toda lattice with a different $\epsilon$ is quite different, and
the solution with at least one $\epsilon_k<0$ has a blow-up at some $t\in\Real$.

\begin{figure}[t!]
\centering
\includegraphics[scale=0.45]{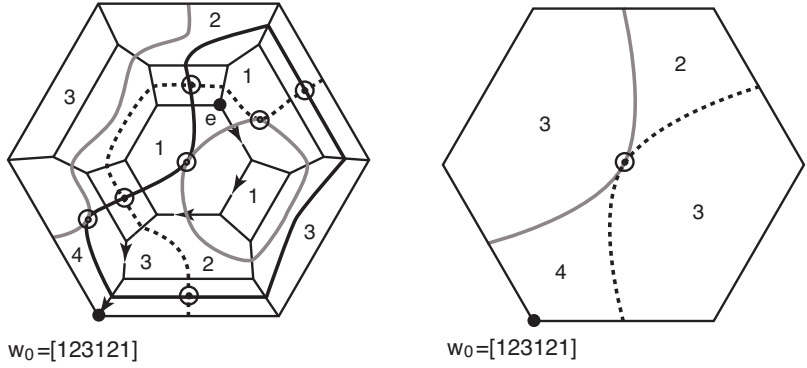}
\caption{The moment polytope $\mathcal{M}_{---}$ for the $\frak{sl}_4(\Real)$ indefinite Toda lattice
(the right figure is the back view of the left one).
The divisors defined by the set of zero points for the $\tau$-functions are shown by 
the dotted curve for $\{\tau_1=0\}$, by the light curve for $\{\tau_2=0\}$, and by the dark one for 
$\{\tau_3=0\}$.
The double circles indicate the divisors with $\{\tau_i=0\}\cap\{\tau_j=0\}$, which are all connected
at the center of the polytope corresponding to the point with $\{\tau_1=\tau_2=\tau_3=0\}$. The numbers in the polytope indicate the number of blow-ups
along the flow. An example of a path from the top vertex $e$ to the bottom vertex $w_0$, the longest
element of $\Sym_4$, is shown by directed edges.}
\label{fig:A3polytope}
\end{figure}

We now consider each edge of the polytope which corresponds to an $\frak{sl}_2(\Real)$ indefinite Toda
lattice, that is, where $g_j\ne 0$ for only one $j$. This edge can be also expressed by
a simple reflection $s_j\in W$.
Since the simple reflection $s_j$ exchanges $\sigma_j$ and $\sigma_{j+1}$, we have an action of $s_j$
on all the signs $\epsilon_k$, $s_j:\epsilon_k\to\epsilon'_k$,
\[
\epsilon'_k=s_j(\epsilon_k)=\left\{ \begin{array}{llll}
\epsilon_k\epsilon_{k-1} \quad &{\rm if} ~~& j=k-1\\[0.4ex]
\epsilon_k\epsilon_{k+1}\quad &{\rm if} ~~& j=k+1\\[0.4ex]
\epsilon_k    \quad &{\rm if}  ~~ &j=k, ~{\rm or}~|j-k|>1
\end{array}\right.
\]
which can be also shown directly from the form of $\tau_k(t)$ in (\ref{sign tau}). This formula can be extended to the indefinite Toda lattice on any real split semisimple Lie algebras, and we have
(see (\ref{general tau}) and Proposition 3.16 in \cite{CK:02}):
\begin{proposition}\label{sign change}
Let $\epsilon_j={\rm sgn}(g_j)$ for $j=1,\ldots,n-1$. Then the Weyl group action on the signs is given by
\[
s_j ~: ~\epsilon_k\quad  \longmapsto \quad \epsilon_k\epsilon_j^{-C_{kj}}\,,
\]
where $(C_{ij})_{1\le i,j\le n-1}$ is the Cartan matrix of $\frak{sl}_n(\Real)$.
\end{proposition}
With this $W$-action on the signs $\epsilon=(\epsilon_1,\ldots,\epsilon_{n-1})$
with $\epsilon_k={\rm sgn}(g_k)$ at each vertex of the polytope, we now define the relation between the
vertices labeled by $w$ and $w'=ws_i$ as follows: Notice that if $\epsilon_i=+$, then $
(\epsilon_1,\cdots,\epsilon_{n-1})$ remains the same under $s_i$-action. Then we write
\[
w\Longrightarrow w' \qquad {\rm with}\quad w'=ws_i\,.
\]
The following definition gives the number of blow-ups in the Toda orbit from the top vertex $e$
to the vertex labeled by $w\in W$:  Choose a reduced expression $w=s_{j_1}\cdots s_{j_k}$, and
consider the sequence of the signs at the orbit given by $w$-action,
\[
\epsilon\, ~\longrightarrow\,~ s_{j_1}\epsilon\,~\longrightarrow ~\,s_{j_2}s_{j_1}\epsilon\,~\longrightarrow\,~ \cdots\,~\longrightarrow ~\,w^{-1}\epsilon\,.
\]
We then define the function $\eta(w,\epsilon)$ as the number of $\to$ which are not of the form
$\Rightarrow$. The number $\eta(w_0,\epsilon)$ for the longest element $w_0$ gives the total
number of blow-ups along the Toda flow in the polytope of $\mathcal{M}_{\epsilon}$.
Whenever $\epsilon=(-,\ldots,-)$, we just denote $\eta(w,\epsilon)=\eta(w)$.
This number $\eta(w,\epsilon)$ does not depend on the choice of the reduced expression
of $w$ (see Corollary 5.2 in \cite{CK:06}). Hence the number of blow-up points along the 
trajectories in the edges of the polytope is independent of the trajectory parametrized by the
reduced expression. In Figure \ref{fig:A2indefinite}, we illustrate the numbers $\eta(w,\epsilon)$
for the $\frak{sl}_3(\Real)$ indefinite Toda lattice.
For example, on $\mathcal{M}_{--}$, we have $\eta(e)=0, \eta(s_1)=\eta(s_2)=\eta(s_1s_2)=\eta(s_2s_1)=1$ and $\eta(s_1s_2s_1)=2$, i.e the total number of blow-ups is 2. We also illustrate this for the $\frak{sl}_4(\Real)$
Toda lattice in Figure \ref{fig:A3polytope}. Along the path shown in this Figure, we have $\eta(e)=0, \eta([2])=\eta([21])=\eta([213])=1,
\eta([2132])=2,\eta([21323])=3$ and $\eta(w_0)=4$, where $[ij\cdots k]=s_is_j\cdots s_k$, and
note $[21323]=[12312]$.

In general, the total number of blow-ups $\eta(w_0,\epsilon)$ depends only
the initial signs $\epsilon=(\epsilon_1,\ldots,\epsilon_{n-1})$ with $\epsilon_i={\rm sgn}(g_i(t))$
for $t\to-\infty$, which is given by $\epsilon_i=\sigma_i\sigma_{i+1}$. Then in the case of $\frak{sl}_n(\Real)$ indefinite
Toda lattice, the number $\eta(w_0,\epsilon)=m(n-m)$ where $m$ is the total number of negative $\sigma_i$'s
(Proposition 3.3 in \cite{KY:98}).
In particular, the maximum number of blow-ups occurs in the case where $\epsilon=(-,\ldots,-)$, and it is given by $[(n+1)/2](n-[(n+1)/2])$. Those numbers $\eta(w_0,\epsilon)$ are related to the polynomials given in \eqref{Fq points} appearing in $\mathbb{F}_q$ points on certain compact groups $K$.

We now introduce polynomials in terms of the numbers $\eta(w,\epsilon)$, which play a key
role for counting the number of blow-ups and give a surprising connection to the rational
cohomology of the maximal compact subgroup $SO(n)$ (Definition 3.1 in \cite{CK:06}).
\begin{definition}\label{p(q) polynomial}
We define a monic polynomial associated to the polytope $\mathcal{M}_{\epsilon}$,
\[
p(q,\epsilon)=(-1)^{l(w_0)}\sum_{w\in W}(-1)^{l(w)}q^{\eta(w,\epsilon)}\,,
\]
where $l(w)$ indicates the length of $w$. Notice that the degree of $p(q,\epsilon)$, denoted by ${\rm deg}(p(q,\epsilon))$,
is the total number of blow-ups, i.e. $\eta(w_0,\epsilon)={\rm deg}(p(q,\epsilon))$. For the case $\epsilon=(-,\ldots,-)$, we simply denote it by $p(q)$.
\end{definition}    
\begin{example}
In the case of the $\frak{sl}_2(\Real)$ Toda lattice, 
\begin{itemize}
\item[(a)] for $\epsilon=(+)$, we have $e\Rightarrow s_1$ which gives $p(q,+)=0$,
\item[(b)] for $\epsilon=(-)$, we have a blow-up between $e$ and $s_1$, hence $p(q,-)=q-1$.
\end{itemize}
Recall from the previous section that the polynomial $p(q)=p(q,-)$ appears in $|SO(2,\mathbb{F}_q)|=q-1$.

In the case of the $\frak{sl}_3(\Real)$ Toda lattice, from Figure \ref{fig:A2indefinite},
\begin{itemize}
\item[(a)] for all the cases of $\epsilon=(\epsilon_1,\epsilon_2)$ except $(-,-)$, we have
$p(q,\epsilon)=0$.
\item[(b)] for $\epsilon=(-,-)$, we have $p(q)=q^2-1$.
\end{itemize}
Note again that the polynomial $p(q)$ appears in $|SO(3,\mathbb{F}_q)|=q(q^2-1)$.

In the case of $\frak{sl}_4(\Real)$, we have, from Figure \ref{fig:A3polytope},
\begin{itemize}
\item[(a)] for all $\epsilon=(\epsilon_1,\epsilon_2,\epsilon_3)$ except $(-,-,-)$, $p(q,\epsilon)=0$.
\item[(b)] for $\epsilon=(-,-,-)$, $p(q)=q^4-2q^2+1=(q^2-1)^2$.
\end{itemize}
Again note that  $|SO(4,\mathbb{F}_q)|=q^2(q^2-1)^2$.
\end{example}
Casian and Kodama prove that the polynomial $p(q)$ for $\mathcal{M}_{\epsilon}$ with
$\epsilon=(-,\ldots,-)$ in Definition \ref{p(q) polynomial} agrees with the polynomial $p(q)$ in 
$|K(\mathbb{F}_q)|$ in (\ref{Fq points}) where $K$ is the maximal compact subgroup
of real split semisimple Lie group $G$ for the Toda lattice (Theorem 6.5 in \cite{CK:06}).

Thus the polynomial $p(q)$ contains all the information on the $\mathbb{F}_q$ points on the
compact subgroup $K$ of $G$, which is also related to the rational cohomology, i.e.
$H^*(K,\mathbb{Q})=H^*(G/B,\mathbb{Q})$ (see (\ref{KG relation})).
Now recall that the integral cohomology of the real flag variety $G/B$ is obtained
by the incidence graph $\mathcal{G}_{G/B}$ in Definition \ref{incidence graph}.
In \cite{CK:06}, Casian and Kodama show that the graph $\mathcal{G}_{G/B}$ can be
obtained from the blow-ups of the Toda flow. They define a graph $\mathcal{G}_{\epsilon}$ associated to the blow-ups as follows:
\begin{definition} 
The graph $\mathcal{G}_{\epsilon}$ consists of vertices labeled by the elements of the Weyl
group $W$ and oriented edges $\Rightarrow$. The edges are defined as follows:
\[
w_1\Rightarrow w_2\quad{\rm iff}\quad \left\{\begin{array}{lllll}
{\rm (a)}~ w_1\le w_2~ ({\rm Bruhat~order})\\[0.5ex]
{\rm (b)}~ l(w_1)=l(w_2)+1 \\[0.5ex]
{\rm (c)}~ \eta(w_1,\epsilon)=\eta(w_2,\epsilon)\\[0.5ex]
{\rm (d)}~w_1^{-1}\epsilon=w_2^{-1}\epsilon
\end{array}\right.
\]
When $\epsilon=(-,\ldots,-)$, we simply denote $\mathcal{G}=\mathcal{G}_{\epsilon}$.
\end{definition}
Then they prove that $\mathcal{G}_{\epsilon}$ with $\epsilon=(-,\ldots,-)$ is equivalent to $\mathcal{G}_{G/B}$ (Theorem 3.5 in \cite{CK:06} which is the main theorem in the paper).
For example, the graph $\mathcal{G}$ associated with Figure \ref{fig:A3polytope}
agrees with the incidence graph $\mathcal{G}_{G/B}$ given in Figure \ref{fig:A3incidence}.
The proof of the equivalence $\mathcal{G}_{G/B}=\mathcal{G}$ contains several technical steps,
which are beyond the scope of this review.



\medskip

\end{document}